\documentclass[aps,amsfonts,amsmath,prd,preprint,nofootinbib,tightenlines]{revtex4}
\usepackage{epsf}

\def\be{\begin{equation}}
\def\ee{\end{equation}}
\def\beq{\begin{equation}}
\def\eeq{\end{equation}}
\def\bea{\begin{eqnarray}}
\def\eea{\end{eqnarray}}
\def\beqa{\begin{eqnarray}}
\def\eeqa{\end{eqnarray}}

\def\obs{{\text{obs}}}

\def\Pchoosem{\left(\begin{array}{c}
P \\ m \\
\end{array} \right)}

\def\Mchoosen{\left(\begin{array}{c}
M \\ n \\
\end{array} \right)}

\def\erfc{{\mathop{\text{erfc}}}}

\begin{document}

\title{Anthropic prediction for a large multi-jump landscape}
\author{Delia Schwartz-Perlov}
\affiliation{Institute of Cosmology, Department of Physics and Astronomy\\
Tufts University, Medford, MA 02155, USA }

\begin{abstract}
The assumption of a flat prior distribution plays a critical role in
the anthropic prediction of the cosmological constant. In a previous
paper we analytically calculated the distribution for the
cosmological constant, including the prior and anthropic selection
effects, in a large toy ``single-jump'' landscape model. We showed
that it is possible for the fractal prior distribution we found to
behave as an effectively flat distribution in a wide class of
landscapes, but only if the single jump size is large enough. We
extend this work here by investigating a large ( $N \sim 10^{500}$)
toy ``multi-jump'' landscape model.  The jump sizes range over three
orders of magnitude and an overall free parameter $c$ determines the
absolute size of the jumps.  We will show that for ``large'' $c$ the
distribution of probabilities of vacua in the anthropic range is
effectively flat, and thus the successful anthropic prediction is
validated. However, we argue that for small $c$, the distribution
may not be smooth.

\end{abstract}

\maketitle

\section{Introduction}

A beautiful feature of inflation is that it is generically eternal
\cite{AV83,Linde86,Starobinsky} giving rise to the ``multiverse''.
Developments in string theory have also led to the complementary
world view \cite{BP,Susskind,AHDK} that the fundamental laws of
physics admit a vast array of possible solutions.  In such models
there are of order $10^{500}$ different solutions/vacua with various
cosmological constants \cite{Douglas,AshokDouglas,DenefDouglas}.
Each vacuum state represents a possible type of bubble universe
governed by its own low-energy laws of physics.

This so-called ``string theory landscape'' of possibilities is
expected to have many high-energy metastable false vacua which can
decay through bubble nucleation\cite{CdL,Parke,BT}. Bubbles of
lower-energy vacuum can nucleate and expand in the high-energy
vacuum background and vice versa\footnote{However, if the
lower-energy vacuum has negative or zero-energy, recycling cannot
take place.  We will call vacua from which new bubbles can nucleate
non-terminal, or recyclable vacua, while those which do not recycle
will be called terminal vacua.}\cite{EWeinberg,recycling}.  This
recycling process will populate the multiverse with bubbles of all
different types nested one within the other.

Most of these bubbles will never be home to observers. For example,
bubbles with large positive cosmological constant do not allow for
structures such as galaxies or atoms to form
\cite{Weinberg87,Linde87}. And bubbles with large negative
cosmological constant collapse long before life has a chance to
evolve.  However, because of the vastness of the landscape, there
will also be many bubbles which do provide a suitable environment in
which life can flourish.  We are not surprised to find ourselves in
such a fertile bubble.

In the context of the multiverse, some physical parameters that were
once thought of as fundamental universal parameters may simply be
``local'' environmental parameters. The most famous example of one
such parameter, is that of the cosmological constant.

The observed value of the cosmological constant $\Lambda$ is about
$120$ orders of magnitude smaller than theoretically
expected\footnote{Here and below we use reduced Planck units,
$M^2_{P} \equiv (8\pi G)^{-1} =1$, where $G$ is Newton's
gravitational constant. The theoretical expectation could be as
``large'' as $10^{-56}$, for example because of supersymmetry.
However, a discrepancy of $56$ orders of magnitude still needs to be
explained.} \be \Lambda_0\sim 10^{-120}. \ee

A very appealing explanation for this observation assumes that
$\Lambda$ is an environmental parameter which has different values
in different parts of the multiverse
\cite{Weinberg87,Linde87,AV95,Efstathiou,MSW,GLV,Bludman,AV05}. The
probability for a randomly picked observer to measure a given value
of $\Lambda$ can then be expressed as \cite{AV95} \be
P_\obs(\Lambda)\propto P(\Lambda)n_\obs(\Lambda), \label{Pobs} \ee
where $P(\Lambda)$ is the prior distribution or volume fraction of
regions with a given value of $\Lambda$ and $n_\obs(\Lambda)$ is the
anthropic/selection factor, which is proportional to the number of
observers that will evolve per unit volume. If we ignore the
variation of other constants, then the density of observers is
roughly proportional to the fraction of matter clustered in large
galaxies, $n_\obs(\Lambda) \propto f_G(\Lambda)$. Using the
Press-Schechter approximation \cite{Press:1973iz} for $f_G(\Lambda)$
we can write \cite{astro-ph/0611573} \be \label{eqn:erfc}
n_\obs(\Lambda) \sim
\erfc\left[\left(\frac{\Lambda}{\Lambda_c}\right)^{1/3}\right] \ee
where we have normalized $n_\obs$ to be 1 for $\Lambda = 0$ and we
have parameterized the anthropic suppression with a value
$\Lambda_c$.  For the parameters used in Refs.
\cite{astro-ph/0611573,astro-ph/0410281}, $\Lambda_c$ is about 10
times the observed value of $\Lambda$, \be \Lambda_c \sim 6\times
10^{-120}. \ee

The prior distribution $P(\Lambda)$ depends on the unknown details
of the fundamental theory and on the dynamics of eternal inflation.
However, it has been argued \cite{AV96,Weinberg96} that it should be
well approximated by a flat distribution, \be P(\Lambda)\approx {\rm
const} \label{flat}, \ee because the window where $n_\obs(\Lambda)$
is substantially different from zero, is vastly less than the
expected Planck scale range of variation of $\Lambda$.  Any smooth
function varying on some large characteristic scale will be nearly
constant within a relatively tiny interval. Thus from Eq.\
(\ref{Pobs}), \be P_\obs(\Lambda)\propto n_\obs(\Lambda).
\label{PnG} \ee

Indeed the observed value of $\Lambda$ is compatible with the
distribution of Eq.\ (\ref{eqn:erfc}). This successful prediction
for $\Lambda$ depends crucially on the assumption of a flat volume
distribution (\ref{flat}).  If, for example, one uses $P(\Lambda)
\propto\Lambda$ instead of (\ref{flat}), the $2\sigma$ prediction
would be $\Lambda/\Lambda_0 <500$, giving no satisfactory
explanation for why $\Lambda$ is so small \cite{Pogosian}.

Given a specific string theory landscape, we would like to be able
to calculate the prior probability $P(\Lambda)$ so that ultimately
we can predict the cosmological constant that we should expect to
observe according to Eq.~(\ref{Pobs}). We will assume $P(\Lambda)$
is given by the relative bubble abundances of different vacua. Since
an eternally inflating multiverse contains an infinite number of
each type of vacuum allowed in the landscape, it is necessary to use
some regularization procedure to compute the prior probability
distribution.  Many such regularization procedures, or probability
measures, have been proposed
\cite{LLM94,Bousso,Bousso:2007nd,Alex,DeSimone:2008bq,Aguirre,Linde07A}.
For a more complete and up to date account see Ref. \cite{Linde07B}
and the references therein. Here we will use the pocket-based
measure introduced in Refs.\ \cite{GSPVW,ELM}. Refs.\ \cite{SPV,SP}
computed prior probabilities\footnote{Strictly speaking bubble
abundances were calculated.} of different vacua in toy models
\cite{BP,AHDK} and found ``staggered'' distributions ranging over
many orders of magnitude.  However, to allow for numerical solution,
Refs.\ \cite{SPV,SP} used models with a relatively small number of
vacua and worked only in a first-order approximation.  See also
Ref's
\cite{Bousso:2007er,Clifton:2007bn,Podolsky:2008du,Podolsky:2007vg}
for related work.

In  Ref.\ \cite{OSP} we considered a toy model in which bubble
abundances were computed as though all changes in $\Lambda$ were by
some fixed amount\footnote{For such a model many degenerate vacua
exist with widely separated cosmological constant.  We therefore
modified the model by artificially perturbing the $\Lambda$ of each
vacuum, producing a smooth \emph{number} distribution.} $\tilde{c}$.
In this simple model, we analytically studied probability
distributions for a realistic number of vacua, $N \sim 10^{500}$. We
found that when $\tilde{c}$ is around 1, there is a smooth
distribution of vacua in the anthropic range, and the anthropic
prediction of Eq.~(\ref{PnG}) applies.  But when $\tilde{c}$ is
smaller by a few orders of magnitude, we found that the $P(\Lambda)$
factor, which favors high $\Lambda$ vacua, is more important than
$n_\obs(\Lambda)$ in Eq.~(\ref{Pobs}). This implies we should expect
to live in a region with large $\Lambda$, and therefore the
anthropic procedure would not explain the observed small value of
$\Lambda$.

In this paper we study a more sophisticated model with a
\emph{range} of jump sizes, parameterized by an overall free
parameter $c$.  We call it the multi-step model (or MS model). For
each value of $c$, the jump sizes range over three orders of
magnitude.

%Thus for a particular value of $c$, we can produce a model with jump
%sizes which range from ``small'', $\mathcal{O}(10^{-3})$ to
%``large'', $\mathcal{O}(1)$ , in the sense of Ref. \cite{OSP}.

We will show that even for a model with a wide range of jump sizes,
the flatness of the prior distribution depends on the free parameter
$c$.  We find that for ``large'' $c$ the prior distribution is flat
but for ``small'' $c$ there will be some staggering.  However it
seems as though the extent of staggering is reduced in the
multi-step model compared to the single-step model (or SS model) of
Ref. \cite{OSP} for comparable small values of $c$ and $\tilde{c}$.

We also study a simpler ``averaged'' version of the model, the
averaged multi-step model (or AMS model), which is more amenable to
analytic evaluation and provides some insight into the general case.

The plan of this paper is as follows:   We will define our model in
section \ref{toymodel}, and arrive at a general expression for the
prior probability distribution in section \ref{probabilities}.  We
also analytically calculate large $c$ limits for the prior
probability distribution and present numerical results and heuristic
arguments for the small $c$ limit in section \ref{probabilities}.

In section \ref{fulldistributionlargec} we will calculate the
distribution for the observed $\Lambda$ in the large $c$ regime.
%,
%and the small $c$ case will be discussed in section
%\ref{fulldistributionsmallc}. We present a qualitative comparison
%between the single-step and multi-step models in section
%\ref{singlemulticompare}.
We end the paper with a discussion in
section \ref{discussion}.

In Appendix \ref{AMS} we will study the averaged multi-step model
and calculate the \emph{prior} probabilities in large and small $c$
regimes. In Appendix \ref{AMSfulldistribution} we will compare
expected \emph{observed} probabilities of vacua reached via $n$ and
$n+1$ jumps for the AMS model. We also include an Appendix in which
we outline the method used to calculate bubble abundances.

\section{The multi-step model} \label{toymodel}

\subsection{Preliminary outline}

We will study a version of the Arkani-Hamed-Dimopolous-Kachru (ADK)
landscape model \cite{AHDK} which has $J$ directions, and $N =
2^{J}$ vacua.  We will choose $J \approx 1600$, so that $N\sim
10^{500}$. Each vacuum in this landscape can be specified by a list
of numbers $\{\eta_1,\ldots,\eta_{2}\}$, where $ \eta_i = \pm 1$,
and the cosmological constant is \be \label{eqn:Lambda}\Lambda =
\bar\Lambda + \frac{1}2\sum_i \eta_i c i\ee

We will take the average cosmological constant $\bar\Lambda$ to be
in the range $(0, c)$.  Each vacuum has $J$ neighbors to which it
can tunnel by bubble nucleation. Each nucleation event results in an
increase or decrease of the cosmological constant\footnote{For the
specific value $c \approx 10^{-3}$, the model has a set of jump
sizes ranging roughly between the ``small'' value of $10^{-3}$ and
the ``large'' value of $1$. The smaller jump sizes are in the regime
where anthropic reasoning breaks down for the single step model
studied in Ref.\ \cite{OSP}, while the larger jump sizes are in the
regime where anthropic reasoning was found to be valid in Ref.\
\cite{OSP}. In the model of Ref.\ \cite{OSP} we considered \be
\Lambda = \bar\Lambda + \frac{1}2\sum_i \eta_i \tilde{c} \ee Thus
all jumps had the same size, $\tilde{c}$.} by $c i$ with $1 \leq i
\leq J = 1600$.

For the model described above vacua exist only with widely separated
cosmological constants, $\bar\Lambda,\bar\Lambda +c$\ldots,
therefore we do not expect any in the anthropic range.  Thus we will
modify the model by artificially perturbing the $\Lambda$ of each
vacuum to produce a smooth number distribution.  Vacua originally
clustered at $\bar\Lambda$ will be spread out over the range from 0
to $c$.  This will cover the anthropic range of vacua with $\Lambda
> 0$, and so if the density of vacua is high enough we will find
some anthropic vacua.  We will not, however, take account of these
perturbations in computing probabilities.

We will only be interested in the vacua near $\Lambda = 0$, which
are those that are in the range of $(0,c)$ before implementing the
perturbation procedure above. One of these will be the so-called
``dominant'' vacuum (defined in Subsection \ref{Dominant vacuum})
with $\Lambda = \Lambda_*$.

%We caution the reader that as we study the multi-jump model we will
%find that due to its complexity we will be forced to abandon our
%analytic analysis and resort to numerical calculations.  However,
%there is a simplified version of the full-blown multi-step model,
%which we will call the ``averaged multi-step'' (or AMS) model, for
%which analytic calculation is possible in two regimes of parameter
%space (small and large $c$).  Thus in Section \ref{AMS} we will
%interrupt our study of the MS model to consider the AMS model which
%we will define later.

\subsection{Nucleation rates in the multi-jump model}
Vacua with $\Lambda\leq 0$ are said to be terminal.  There are no
transitions out of them.  Vacua with $\Lambda>0$ are recyclable.  If
$j$ labels such a vacuum, it may be possible to nucleate bubbles of
a new vacuum, say $i$, inside vacuum $j$. The transition rate
$\kappa_{ij}$ for this process is defined as the probability per
unit time for an observer who is currently in vacuum $j$ to find
herself in vacuum $i$. Using the logarithm of the scale factor as
our time variable, \be
\kappa_{ij}=\Gamma_{ij}\frac{4\pi}{3}H_j^{-4}, \label{kappa} \ee
where $\Gamma_{ij}$ is the bubble nucleation rate per unit physical
spacetime volume (defined in next subsection and the same as
$\lambda_{ij}$ in \cite{GSPVW}) and \be H_j = (\Lambda_j/3)^{1/2}
\label{hubble} \ee is the expansion rate in vacuum $j$.

Transitions between neighboring vacua, which change one of the
integers $\eta_a$ by $\pm 2$ can occur through bubble nucleation.
The bubbles are bounded by thin branes, with tension $\tau_a$.
Transitions with multiple brane nucleation, in which several
$\eta_i$ are changed at once, are likely to be strongly suppressed
\cite{Megevand}, and we shall disregard them here.

The bubble nucleation rate $\Gamma_{ij}$ per unit spacetime volume
can be expressed as \cite{CdL} \be \Gamma_{ij}=A_{ij} \exp^{-B_{ij}}
\label{Gamma} \ee with \beq B_{ij}=I_{ij}-S_j \label{Bij} \eeq Here,
$I_{ij}$ is the Coleman-DeLuccia instanton action and \beq
S_j=-\frac{8\pi^2}{H_j^2} \label{Sj} \eeq is the background
Euclidean action of de Sitter space with expansion rate $H_j$.

In the relevant case of a thin-wall bubble, the instanton action
$I_{ij}$ has been calculated in Refs.~\cite{CdL,BT}. It depends on
the values of $\Lambda$ inside and outside the bubble and on the
brane tension $\tau$.

Let us first consider a bubble which changes one $\eta_a$ from
$\eta_a=+1$ to $\eta_a=-1$.  The resulting change in the
cosmological constant is given by \be |\Delta\Lambda_a|=c a
\label{ADKDeltaLambda} \ee and the exponent in the tunneling rate
(\ref{Gamma}) can be expressed as \be B_{a\downarrow} =
B_{a\downarrow}^{flatspace} r(x,y). \label{ADKBdown} \ee
$B_{a\downarrow}^{flatspace}$ is the flat space bounce action, \be
B_{a\downarrow}^{flatspace}= \frac{27
\pi^2}{2}\frac{\tau_a^4}{|\Delta \Lambda_a|^3}.
\label{equn:flatB}\ee

The gravitational correction factor $r(x,y)$ is given by
\cite{Parke} \be r(x,y) = \frac{2[(1+x
y)-(1+2xy+x^2)^{\frac{1}{2}}]}{x^2(y^2-1)(1+2xy+x^2)^{\frac{1}{2}}}
\label{gravfactor} \ee  with the dimensionless parameters \be
x\equiv \frac{3\tau_a^2}{4|\Delta\Lambda_a|} \label{ADKx}\ee and \be
y\equiv \frac{2\Lambda}{|\Delta\Lambda_a|}-1, \label{ADKy} \ee where
$\Lambda$ is the background value prior to nucleation.
%\frac{\Lambda_{n_j}+\Lambda_{n_j-1}}{\Lambda_{n_j}-\Lambda_{n_j-1}}=
%\frac{\Lambda_{n_j}+\Lambda_{n_j-1}}{\Delta \Lambda}.
%\ee

The brane tension $\tau_i$ enters the tunneling exponent through the
dimensionless parameter $x$ (\ref{ADKx}). We will assume that the
potentials in our model have the same shape but differ by an overall
factor so that $V_i(\phi) = g_i^2 \mathcal{V}(\phi)$. This gives
rise to the set of different $\Delta\Lambda_i$ and $\tau_i$.
However, in this realization of the class of ADK models, the ratio
of $\tau_i^2/{\Delta\Lambda_i}$ remains a constant for each
$V_i(\phi)$.  The results we present will correspond to choosing
this constant to be $1$, which is equivalent to having $x=3/4$.
Although this choice is somewhat ad hoc\footnote{According to Ref.
\cite{Ceresole:2006iq}, the BPS maximum bound on $\tau_i$ enforces
$x<1$.  One can show that small values of $x$ lead to enhancement of
nucleation rates, which tend to smooth out the distribution.  Thus
$x=3/4$ is a reliable choice because it doesn't ``wash out''
potential problems with staggering.}, we do not expect the main
qualitative feature of the results to depend on the precise value of
$x$

\begin{figure}
\begin{center}
\leavevmode\epsfxsize=5in\epsfbox{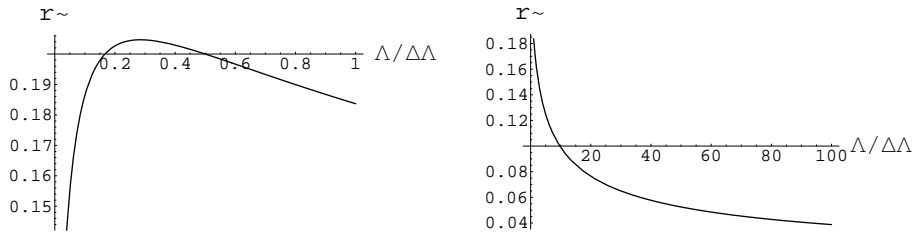}
\end{center}
\caption{The factor $\tilde{r}$ as a function of
$\Lambda/\Delta\Lambda$ for the ranges $0<\Lambda/\Delta\Lambda<1$
and $1<\Lambda/\Delta\Lambda<100$.} \label{rtilde}
\end{figure}

For later convenience, we can rewrite Eq.\ (\ref{ADKBdown}) as

\be B_{a\downarrow}= \frac{24 \pi^2}{\Lambda_a} \tilde{r}(x,y)
\label{newflatB}\ee

where we define (see Fig.~\ref{rtilde}) \be \tilde{r}(x,y) \equiv
\frac{[(1+x
y)-(1+2xy+x^2)^{\frac{1}{2}}]}{(y-1)(1+2xy+x^2)^{\frac{1}{2}}}
\label{newgravfactor} \ee

The prefactors $A_{ij}$ in (\ref{Gamma}) can be
estimated\footnote{$A_{ij}$ is very hard to compute, but using
dimensional arguments \cite{Mukhanov} $A_{ij} \sim
\mathcal{O}(H_j^4) \sim 1$ for $\Lambda_j \sim 1$. For
$\Lambda_j<<1$, we expect this result to hold because the tunneling
rate remains finite in the limit $\Lambda_j\rightarrow 0$,
$|\Delta\Lambda_a| \sim 1$.} as \cite{Jaume} \beq A_{ij} \sim 1
.\eeq

If the vacuum $\{\eta_1...\eta_{a-1},\eta_a-2,\eta_{a+1}....\}$
still has a positive energy density, then an upward transition from
$\{\eta_1...\eta_{a-1},\eta_a-2,\eta_{a+1}....\}$ to
$\{\eta_1...\eta_{a-1},\eta_a,\eta_{a+1}....\}$ is also possible.
The corresponding transition rate is characterized by the same
instanton action and the same prefactor \cite{EWeinberg}, and it
follows from Eqs. (\ref{Gamma}), (\ref{Bij}) and (\ref{hubble}) that
the upward and downward nucleation rates are related by \be
\Gamma_{ji} = \Gamma_{ij} \exp\left[-24 \pi^2
\left(\frac{1}{\Lambda_{i}}-\frac{1}{\Lambda_{j}}\right)\right]
\label{updown} \ee where $\Lambda_j>\Lambda_i$.  As expected the
transition rate from $\eta_a=-1$ up to $\eta_{a}=+1$ is suppressed
relative to that from $\eta_{a}=+1$ down to $\eta_{a}=-1$. The
closer we are to $\Lambda_i=0$, the more suppressed are the upward
transitions $i\to j$ relative to the downward ones.

%Transition rates from a given vacuum $j$ to different states $i$ are
%related by \beq \Gamma_{ij}\propto \exp(-I_{ij}). \label{GammaI}
%\eeq As a rule of thumb, \beq I_{ij}\sim -\Lambda_{max}^{-1},
%\label{ILambda} \eeq where $\Lambda_{max}$ is the larger of
%$\Lambda_i$ and $\Lambda_j$. It follows from
%(\ref{GammaI}),(\ref{ILambda}) that upward transitions from a given
%site are more enhanced to lower-energy vacua.

We will now approximate the dependence of the tunneling exponent
$B_{a\downarrow}$ on the parameters of the model in the following
three regimes:

\begin{enumerate}
  \item In the limit, $\Lambda =|\Delta\Lambda_a|$, $y = 1$, \beq
\tilde{r}(3/4,y=1)\approx \frac{9}{49} \approx
\frac{1}{5}.\label{rapproxy1} \eeq
  \item For $\Lambda\gg |\Delta\Lambda_a|$, we have $y\gg1$, and Eq.\
(\ref{newgravfactor}) gives \beq \tilde{r}(x,y \gg 1)\approx
\frac{1}{\sqrt{2}}\sqrt{\frac{x}{y}} \approx
\frac{\sqrt{3}}{4}\sqrt{\frac{\Delta\Lambda}{\Lambda}}
\label{rapproxlargey}\eeq where we used $x=3/4$ in the last step.
  \item For $\Lambda\ll |\Delta\Lambda_a|$, we have $y\approx-1$, and Eq.\
(\ref{newgravfactor}) gives \beq \tilde{r}(x,y \approx -1)\approx
9\frac{\Lambda}{\Delta\Lambda} \label{rapproxyminus1}\eeq
\end{enumerate}

%For later use we note that for $y \gtrsim 3$

%\be \tilde{r}(x,y)<
%\frac{\sqrt{3}}{4}\sqrt{\frac{\Delta\Lambda}{\Lambda}} \equiv
%\tilde{r}_{large} \label{rlarge} \ee and

%\be \tilde{r}(x,y) >
%\frac{\sqrt{3}}{8}\sqrt{\frac{\Delta\Lambda}{\Lambda}} \equiv
%\tilde{r}_{small}\label{rsmall}.\ee

For the first two regimes above, $\tilde{r}(x,y)<1$, thus the
inclusion of gravity decreases the tunneling exponent causing an
enhancement of the nucleation rate\footnote{There are other regimes
in which gravity causes suppression of the nucleation rate
\cite{CdL}. However, in this paper, all the transitions which will
enter in the comparison of probabilities will have $\Lambda\geq
|\Delta\Lambda_a|$ and thus are in the graviationally enhanced
regime.}. We note that the use of the semi-classical approximation
is justified when the tunneling action is large enough:
$B_{a\downarrow}\gg 1$.

\subsection{Definition of the dominant vacuum}\label{Dominant vacuum}

If we define the total down-tunneling rate for a vacuum $j$, \be D_j
= \sum_{\Lambda_i < \Lambda_j}\kappa_{ij}\, \label{dominant}\ee

then the dominant vacuum, referred to as vacuum $*$, is defined as
that recyclable vacuum whose $D_j$ is the smallest.  We will call
$D_j$ of the dominant vacuum $D_*$. Since bubble nucleation rates
are suppressed in low-energy vacua, we expect $\Lambda_*$ to be
fairly small, however we would not expect it to be so small as to be
in the anthropic range.  This can be understood as follows:  from
Eq.'s (\ref{newflatB}) and (\ref{rapproxyminus1}) we see that the
transition rates become independent of $\Lambda$ once we are in the
regime $\Lambda \ll \Delta\Lambda_j $.  For example, when $\Lambda
\sim 10^{-3}c$ (still vastly bigger than the anthropic range $\sim
\mathcal{O}(10^{-119})$ for any reasonable value of $c$) the
transition rates only depend on $\Delta\Lambda_j$, so it is possible
for vacuum $y$ with $\Lambda_x < \Lambda_y <\Lambda$ to have a
smaller transition rate than vacuum $x$.

In Bousso-Polchinski and Arkani-Hamed-Dimopolous-Kachru type
landscapes \cite{BP,AHDK} it can be shown that this vacuum will have
no downward transitions to vacua with positive $\Lambda$. To see
that this is true, imagine that in some direction $\Lambda_*$ can
jump downward to $\Lambda_{\alpha}>0$. Now if we compare
$D_{\alpha}$ to $D_*$ we see that each term contributing to
$D_{\alpha}$ is less than the corresponding term (i.e., the
transition rate in the same direction) in $D_*$ because
$\Lambda_{\alpha}<\Lambda_*$ and jump sizes in the same direction
are the same.  This implies $D_{\alpha}<D_*$ which contradicts our
definition of $D_*$ as the vacuum with the smallest sum of downward
transition rates. Thus for each $\kappa_{ij}$ in $D_*$, $\Lambda <
\Delta\Lambda_j $.

Recall that when $\Lambda \ll \Delta\Lambda_j $, the transition
rates $\kappa_{ij}$ only depend on $\Delta\Lambda$ as seen from
Eq.'s (\ref{newflatB}) and (\ref{rapproxyminus1}).  In this regime,
the larger $\Delta\Lambda$ the larger $\kappa_{ij}$.  Thus we expect
the configuration of the dominant vacuum to have $\eta_i=-1$ for the
largest $\Delta\Lambda_i$'s, so that these large transitions are
excluded from the sum in Eq. (\ref{dominant}).

Combining this insight with the fact that $\Lambda_*$ should be
small, we expect the dominant vacuum to have a configuration with
$\eta_i=\{+...+-...-\}$ where the first $P = 0.71 J \approx 1132$
coordinates are $+$'s and the last $M = 0.29 J \approx 468$
coordinates are $-$'s.  These numbers are calculated by setting the
maximum number of the largest $\eta_i$ to be negative, whilst still
ensuring that $\Lambda>0$.

%For the interested reader, more details regarding the expected
%configuration of the dominant vacuum are discussed in Appendix
%\ref{Dominantvacuum}.

\subsection{Distribution of vacua}

For a typical realization of the model with $J=1600$, the dominant
vacuum has $+$ coordinates in roughly the first $P \approx 1132$
directions, and $-$ coordinates in the last $M \approx 468$
directions. We also note that the average jump size of the plus and
minus coordinates respectively are $\Delta\bar{\Lambda}_P \approx
566 c$ and $\Delta\bar{\Lambda}_M \approx 1366 c$. Thus the other
vacua with $0<\Lambda<c$ are reached from the dominant vacuum by
taking a different number of up jumps and down jumps.

What happens when we take $n$ jumps up from the dominant vacuum?  We
get a number distribution of vacua which is peaked around \be
\Lambda_{peak}^{(up)}= \Lambda_*+n \Delta\bar{\Lambda}_M\ee with a
standard deviation which we denote $\sigma^{(up)}_n$.  We find
\be\sigma^{(up)}_n \approx \sqrt{n}135c.\ee

What happens when we take $m$ jumps down from one of the vacua in
the above distribution? We get another number distribution of vacua
which is peaked around \be \Lambda_{peak}^{(down)}= \Lambda_{start}
- m \Delta\bar{\Lambda}_P\ee with a standard deviation which we
denote $\sigma^{(down)}_m$. We find

\be \sigma^{(down)}_m \approx \sqrt{m}327c.\ee

So overall for a ``path'' with $n$ up and $m$ down jumps, we get a
total distribution which is peaked at \be \Lambda_{peak}^{(total)}=
\Lambda_*+n \Delta\bar{\Lambda}_M- m \Delta\bar{\Lambda}_P\ee with
\be\sigma^{(total)}_{n,m}
=\sqrt{{\sigma^{(up)}_n}^2+{\sigma^{(down)}_m}^2}\label{sigmatotal}.\ee

Ultimately, when we come to calculating the probabilities associated
with vacua in our landscape, we will only be interested in those
which are very close to $\Lambda=0$ (which means in our model those
that have cosmological constants in the interval $0<\Lambda<c$).
Thus we are especially interested in the special case of

\be \Lambda_{peak}^{(total)} \sim 0\ee which holds when

\be m \approx n
\frac{\Delta\bar{\Lambda}_M}{\Delta\bar{\Lambda}_P}\sim 2.4n\ee

In other words, if you go up $n$ steps, and down $m$ steps, the
resultant distribution will have its peak around $\Lambda=0$ if
$m=\alpha n$ with $\alpha \equiv
\frac{\Delta\bar{\Lambda}_M}{\Delta\bar{\Lambda}_P} \approx 2.4$.

\begin{figure}
\begin{center}
\leavevmode\epsfxsize=5in\epsfbox{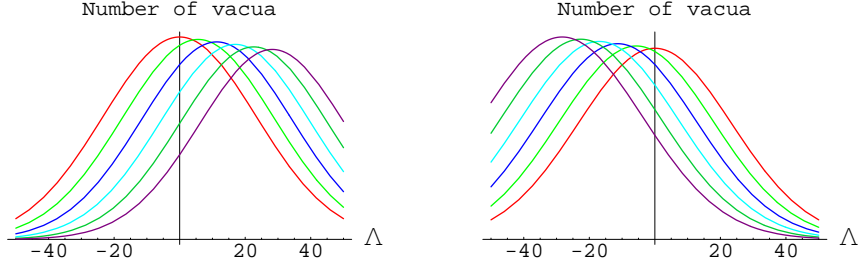}
\end{center}
\caption{On the left we plot six normal distributions with mean
$(i-1)566c$ and standard deviation $\sigma^{(total)}_{n,m}$ (see Eq.
(\ref{sigmatotal})) with $n=20$ and $m=\alpha n-i+1$, and here
$1<i<6$. On the right we plot six normal distributions with mean
$(1-i)566c$ and standard deviation $\sigma^{(total)}_{n,m}$ with
$n=20$ and $m=\alpha n+i-1$, and again $1<i<6$.  We take
$c=10^{-2}$. The heights of these distributions are schematic - as
we take more downward steps from a starting point, the number
distribution grows, as indicated by the increasing height of the
distributions from right to left in each plot.}\label{spread}
\end{figure}

What happens if , for a given $n$ we go down $\alpha n-1$ or $\alpha
n+1$ steps?  The peaks of these distributions will no longer be at
$\Lambda=0$, but if they are broad enough then they may still
``straddle'' the regime of interest $0<\Lambda<c$ with a
significantly high number density (see Fig.\ref{spread}).  We will
consider all vacua within a $1\sigma$ spread to be statistically
significant. It turns out that for $n \approx 20$, the number
distribution of vacua which go down $m$ jumps with $\alpha
n-5\lesssim m \lesssim \alpha n+5 $, straddle the range
$0<\Lambda<c$ within $1 \sigma$ of their respective peaks.  This
means that there is a statistically meaningful number of vacua in
$0<\Lambda<c$ which are reached via all these different ``levels''
of trajectories.

We will classify all vacua in $0<\Lambda<c$ by two parameters: $n$
and $m$, where $n$ is the minimum number of up jumps required to
reach a given vacuum from the dominant vacuum and $m$ is the number
of down jumps. We will call $nm$ the level of the vacuum. Thus a
vacuum of level $nm$ differs from the dominant vacuum in $n+m$
coordinates, $n$ of which are $+$ where the dominant vacuum had $-$,
and another $m$ vice versa.

The total number of vacua of level $nm$ is thus

\be N_{nm} = N^{(up)}_n \times N^{(down)}_{m}\label{eqn:Nn}\ee where

\be N^{(up)}_n = \Mchoosen =\frac{M!}{n!(M-n)!}\label{eqn:Nupn} \ee

and

\be N^{(down)}_{m} = \Pchoosem
=\frac{P!}{m!(P-m)!}\label{eqn:Ndownn} \ee

We approximate these to be smeared over a range $\sigma_{nm}$, where
$\sigma_{nm} \equiv \sigma^{(total)}_{n,m}$, so their density is \be
\rho_{nm} = N_{nm}/\sigma_{nm} \equiv 1/\Delta_{nm}\,. \ee

%For the AMS model studied in Appendix \ref{AMS}, we take
%$\sigma_{nm} \equiv c$.

The likelihood that there is no vacuum in a range of size $x$ is
$\exp(-\rho_{nm} x)$, thus the median $\Lambda$ of the
lowest-$\Lambda$ vacuum is \be \label{eqn:Lambdan} \Lambda_{nm} =
(\ln2)/\rho_{nm} = \sigma_{nm}(\ln2)/N_{nm} \ee We will assume the
lowest-$\Lambda$ vacuum is at this median position. Above the
lowest-$\Lambda$ vacuum of level $nm$, there are $N_{nm}-1$ more
with higher $\Lambda$, with the typical interval in $\Lambda$ being
$\Delta_{nm}$.  For a typical realization it is sufficient to take
these vacua as evenly spaced, so that they are at \be
\Lambda_{nm,\ell} =\Lambda_{nm}+(\ell-1)\Delta_{nm}
\label{eqn:lambdani} \ee where $1 \leq \ell \leq N_{nm}$.

This model is qualitatively different from the single step model
studied in \cite{OSP}.  In the single step model, every vacuum in
the range $(0,c)$ that was reached via $n$ up and \emph{necessarily}
$n$ down jumps (the so-called level $n$ vacua) would have the same
number density $\rho_n$ (and also the same prior probability $P_n$).
In fact, there is only one way to achieve a density of $\rho_n$ and
that is to go up and down $n$ jumps.  The next level would have
$\rho_{n+1} \approx 500 \rho_n$.

\begin{figure}
\begin{center}
\leavevmode\epsfxsize=5in\epsfbox{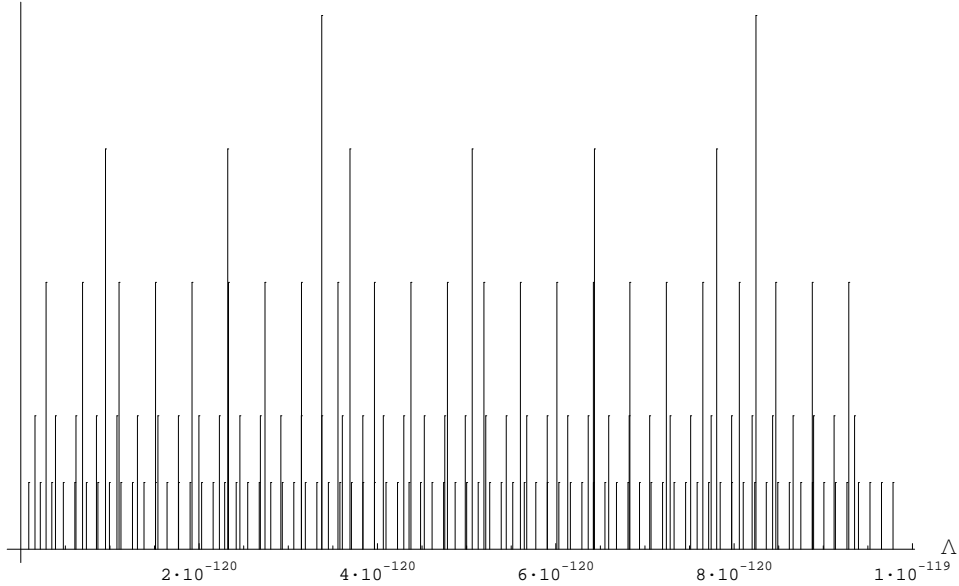}
\end{center}
\caption{Each line represents where we expect to find vacua.  The
highest line is at $\Lambda_{nm} \sim 3.4 \times 10^{-120}$ with
$n=19, m = \alpha n +4$. The next set of highest lines start at
$\Lambda_{nm} \sim 9.5 \times 10^{-121}$} and then are spaced by the
corresponding $\Delta_{nm}$ with $n=20, m=\alpha n+1$. Also shown
are two subsets of vacua reached via $n=21, m=\alpha n-2$ and $n=19,
m=\alpha n+5$.  Finally, the last and most dense set shown is for
the first few vacua reached via $n=22, m=\alpha n-5$. The height of
the lines has no meaning here.  We simply want to show how different
level $nm$ paths result in sets of vacua with similar densities. We
have set $c=0.01$ here.\label{picketc0point01}
\end{figure}

On the other hand, in the multi-step model, there are multiple
classes of trajectories which can result in approximately the same
densities $\rho_{nm}$.  For example for $n=20$ and $m=\alpha n+1
\sim 2.4 n+1$, $\rho_{nm} \sim 7.3 \times 10^{119}$ (and
$\Lambda_{nm} \sim 9.5 \times 10^{-121}$) while for $n=21$ and
$m=\alpha n-2$ we find $\rho_{nm} \sim 2.4 \times 10^{120}$ (and
$\Lambda_{nm} \sim 2.8 \times 10^{-121}$).  In Table \ref{tableprob}
we display where we expect the first vacuum $\Lambda_{nm}$ to lie
for a range of values of $n$ and $m$. In Fig. \ref{picketc0point01}
we depict where we expect some of these $\Lambda_{nm}$ vacua to lie,
along with second and several subsequent vacua for the specific
level according to Eq. (\ref{eqn:lambdani}).

\begin{table}
\centering
\begin{tabular}{|l|c|c|c|c|}
\hline  m & n=19 & n=20 & n=21 & n=22 \\
\hline
      $\alpha n+5$ & $\mathbf{1.6 \times 10^{-121}}$ &  $5.0 \times 10^{-126}$ &  $1.9 \times 10^{-130}$ & $8.3 \times 10^{-135}$  \\
      $\alpha n+4$ & $\mathbf{3.4 \times 10^{-120}}$ &  $1.0 \times 10^{-124}$ &  $3.7 \times 10^{-129}$ & $1.5 \times 10^{-133}$ \\
      $\alpha n+3$ & $7.3 \times 10^{-119}$ &  $2.1 \times 10^{-123}$ &  $7.2 \times 10^{-128}$ & $2.9 \times 10^{-132}$ \\
      $\alpha n+2$ & $1.6 \times 10^{-117}$ &  $\mathbf{4.4 \times 10^{-122}}$ &  $1.4 \times 10^{-126}$ & $5.5 \times 10^{-131}$ \\
      $\alpha n+1$ & $3.6 \times 10^{-116}$ &  $\mathbf{9.5 \times 10^{-121}}$ &  $2.9 \times 10^{-125}$ & $1.1 \times 10^{-129}$ \\
      $\alpha n  $ & $8.4 \times 10^{-115}$ &  $2.1 \times 10^{-119}$ &  $6.1 \times 10^{-124}$ & $2.1 \times 10^{-128}$ \\
      $\alpha n-1$ & $2.0 \times 10^{-113}$ &  $4.7 \times 10^{-118}$ &  $\mathbf{1.3 \times 10^{-122}}$ & $4.3 \times 10^{-127}$ \\
      $\alpha n-2$ & $4.8 \times 10^{-112}$ &  $1.1 \times 10^{-116}$ &  $\mathbf{2.8 \times 10^{-121}}$ & $9.0 \times 10^{-126}$ \\
      $\alpha n-3$ & $1.2 \times 10^{-110}$ &  $2.5 \times 10^{-115}$ &  $6.3 \times 10^{-120}$ & $1.9 \times 10^{-124}$ \\
      $\alpha n-4$ & $3.0 \times 10^{-109}$ &  $6.0 \times 10^{-114}$ &  $1.4 \times 10^{-118}$ & $\mathbf{4.1 \times 10^{-123}}$ \\
      $\alpha n-5$ & $7.8 \times 10^{-108}$ &  $1.5 \times 10^{-112}$ &  $3.3 \times 10^{-117}$ & $\mathbf{9.0 \times 10^{-122}}$ \\
 \hline
 \end{tabular}
 \caption{$\Lambda_{nm}$ for various values of $n$ and
$m$ with $c=0.01$. We have highlighted some values to draw attention
to the fact that paths reached via different levels can have almost
the same $\Lambda_{mn}$ to within a factor of roughly 3 (and
similarly for $\rho_{nm}$ and $\Delta_{nm}$):  In particular when
$(n,m)=(19, 2.4n +4)$, $\Lambda_{nm} \sim 3.4 \times 10^{-120}$.
Then when $(n,m)=(20, \alpha n+1)$ $\Lambda_{nm} \sim 9.5 \times
10^{-121}$ and for $(n,m)=(21, \alpha n-2)$, $\Lambda_{nm} \sim 2.8
\times 10^{-121}$.
 Also for  $(n,m)=(22, \alpha n-5)$, $\Lambda_{nm} \sim 9 \times
10^{-122}$.  As we went ``diagonally'' down, $\Lambda_{nm}$
decreased by a factor of $\sim 3$.  The same pattern can be seen for
the ``diagonal'' immediately above the one just discussed, starting
with $(n,m)=(19,\alpha n+5)$ and $\Lambda_{nm} \sim 1.6 \times
10^{-121}$.}\label{tableprob}
\end{table}

%\begin{figure}
%\begin{center}
%\leavevmode\epsfxsize=5in\epsfbox{ADKMJpicket.eps}
%\end{center}
%\caption{Each line represents where we expect to find vacua.  The
%highest line is at $\Lambda_{nm} \sim 1.9 \times 10^{-119}$ with
%$n=19, m = 2.4n +5$. The next set of highest lines start at
%$\Lambda_{nm} \sim 5.4 \times 10^{-120}$} and then are spaced by the
%corresponding $\Delta_{nm}$ with $n=20, m=2.4n+2$. Also shown are a
%subset of the vacua reached via $n=21, m=2.4n-1$, $n=22, m=2.4n-4$
%and the most dense set shown is for the first few vacua reached via
%$n=20, m=2.4n+3$. The height of the lines has no meaning here.  We
%simply want to show how different level $nm$ paths result in sets of
%vacua with similar densities. We have set $c=1$
%here.\label{ADKMJpicket}
%\end{figure}

In principle, vacua reached via these different classes will have
different probabilities.  Also, there can even be a spread of
probabilities amongst level $nm$ vacua\footnote{In contrast to the
single-step model which assigns a unique prior probability to every
level $n$ vacuum.}, depending on the paths details. This spread is
minimized for large jump sizes.

\section{Probabilities}\label{probabilities}

We now use the formalism of Refs.\ \cite{GSPVW,SPV} outlined in
Appendix\ \ref{abundances} to calculate the relative abundances of
different vacua in our toy model.

The relative abundance of each vacuum $\beta$ is given by a sum over
all chains that connect it to the dominant vacuum, Eq.\
(\ref{eqn:pl}).  The minimum number of transitions in such a chain
is roughly $n+m$.  Longer chains can be formed by jumping one way
and then later the opposite way in the same direction. These chains
will have extra suppression factors because of the extra jumps, but
it remains a possibility that judiciously chosen ``extra'' jumps
might enhance the probabilities enough to over compensate for these
extra suppression factors.  We will take the possibility of
``extra'' jumps into account when we calculate our numerical
results. For now we shall we forget about them and include only
minimum-length chains. Furthermore, the paths that maximize the
bubble abundances are those that entail first making all the up
jumps and then following with a sequence of down jumps. The reason
is that the up-jump suppression factor, Eq.\ (\ref{eqn:updown}), is
least when the starting $\Lambda$ for the jump is highest.  Thus it
is best not to jump down until all necessary up-jumps have been
made.

Also, it can be shown that the path that contributes the most to the
bubble abundance factor of a specific vacuum entails taking up-jumps
in order of decreasing jump size while the down-jumps should be
taken in increasing step sizes.  This is because, jumps taken at
lower $\Lambda's$ are more suppressed than those taken at higher
$\Lambda's$, so the best strategy is to get to high-$\Lambda's$ as
soon as possible.

%\textbf{Is it ever better to make a large extra up jump followed by
%the necessary jumps?  I think extra up jumps can be included to any
%trajectory, and helps to lessen the difference in probabilities
%across the board, for ``large'' and ``small'' resultant vacua.  But
%the issue is not exactly which path is the best. Rather that if the
%large jumps ar more prob, then there are many high prob paths with
%similar probabilities...}

%So we only consider the contribution of paths which consist of
%making all upward jumps first ordered from largest to smallest jumps
%and then following with downward jumps arranged in the reverse
%order.

We can reorganize Eq.\ (\ref{eqn:pl}), \be \label{eqn:pl2} p_\beta =
\sum \frac{\kappa_{\beta
a}}{D_a-D_*}\frac{\kappa_{ab}}{D_b-D_*}\cdots
\frac{\kappa_{rs}}{D_{s}-D_*} \kappa_{st}
\frac{\kappa_{tu}}{D_t-D_*}\cdots \frac{\kappa_{z*}}{D_z-D_*} \ee

The transition rates to the right of the factor $\kappa_{st}$ in Eq.
(\ref{eqn:pl}) are upward rates, and those to the left are downward
rates. $\kappa_{st}$ represents the first downward jump after having
made $n$ upward jumps from the dominant vacuum.

We will approximate $D_*\ll D_j$, since the transition rates are
suppressed for low $\Lambda$ vacua. Now consider the denominator
$D_j$.

\be D_j \propto \exp \left[\frac{-24
\pi^2}{\Lambda_j}\tilde{r}_{jmin}\right] + \cdots + \exp
\left[\frac{-24 \pi^2}{\Lambda_j}\tilde{r}_{jmax}\right]\ee where
$\tilde{r}_{jmin}$ is calculated from Eq.s\ (\ref{newgravfactor})
and (\ref{ADKy}) with $\Delta\Lambda_a$ in the smallest possible
$\Delta\Lambda$ direction \footnote{Consider vacuum $\Lambda_j$ and
let us define $J_j$ to be equal to the number of $+$'s in the
configuration describing vacuum $\Lambda_j$.  There will be $J_j$
different ways to transition down out of $\Lambda_j$ to a lower
$\Lambda$ neighboring vacuum.  Each one of these possible
transitions will have a different factor of $\tilde{r}$ in the
relevant transition rate. This factor is least when a jump occurs in
the smallest $\Delta\Lambda$ direction, and we call it
$\tilde{r}_{jmin}$.}. We expect the smallest possible direction
$\Delta\Lambda \sim 1 c $.

Similarly $\tilde{r}_{jmax}$ is calculated with $\Delta\Lambda_a$ in
the largest possible $\Delta\Lambda$ direction. All other terms in
between have the same form with the $\tilde{r}_j$ factors taking on
all relevant values as $\Delta\Lambda_a$ increases from the smallest
to largest relevant values.

Using Eq.\ (\ref{eqn:updown}), the product of all up-jump
suppression factors is \be \label{eqn:suppression} S =
\left(\frac{\Lambda_{max}}{\Lambda_*}\right)^2\exp \left[-24
\pi^2\left(\frac{1}{\Lambda_*}-\frac{1}{\Lambda_{max}}\right)\right]
\ee where $\Lambda_{max}$ is the maximum $\Lambda$ reached for a
given path.

The first term in the exponent above and the factor $\Lambda_*^{-2}$
will contribute to every vacuum reached via some path from the
dominant vacuum and will thus disappear when we take ratios of
probabilities. We define  \be \tilde{S} \equiv \exp \left[24
\pi^2\left(\frac{1}{\Lambda_{max}}\right)\right]\ee and call it an
up-jump ``enhancement factor''.

Consider the last term in Eq.\ (\ref{eqn:pl2}) which results from
the first upward jump \be \frac{\kappa_{z*}}{D_z-D_*}
=\frac{\kappa_{*z}}{D_z-D_*}
\left(\frac{\Lambda_{z}}{\Lambda_*}\right)^2\exp \left[-24
\pi^2\left(\frac{1}{\Lambda_*}-\frac{1}{\Lambda_z}\right)\right]
\approx \frac{\kappa_{*z}}{D_z}
\left(\frac{\Lambda_{z}}{\Lambda_*}\right)^2\exp \left[-24
\pi^2\left(\frac{1}{\Lambda_*}-\frac{1}{\Lambda_z}\right)\right]\ee

Similarly, the product of $\kappa/D's$ for all the upward jumps can
be written in terms of downward transition rates and an overall
suppression factor as

\be \frac{\kappa_{tu}}{D_t-D_*}\cdots \frac{\kappa_{z*}}{D_z-D_*}
\approx \frac{\kappa_{ut}}{D_t}\cdots
\frac{\kappa_{*z}}{D_z}\left(\frac{\Lambda_{max}}{\Lambda_*}\right)^2\exp
\left[-24
\pi^2\left(\frac{1}{\Lambda_*}-\frac{1}{\Lambda_{max}}\right)\right].\ee

Now we see that we can call $\kappa_{ij}/D_j$ a branching ratio,
since it is the fraction of one possible downward transition rate
from vacuum $\Lambda_j$ divided by the sum of all the possible
downward transition rates from the same vacuum.

The down-jumps are similar, except that the first jump down from
$\Lambda_{max}$ doesn't have a factor of $D_t$ in the denominator
(see Eq.\ (\ref{eqn:pl2})), giving \be \frac{\kappa_{\beta
a}}{D_a-D_*}\cdots \frac{\kappa_{rs}}{D_s-D_*} \kappa_{st} \approx
\frac{\kappa_{\beta a}}{D_a}\frac{\kappa_{ab}}{D_b}\cdots
\frac{\kappa_{rs}}{D_{s}} \kappa_{st}\label{noDj}\ee

\subsection{The prior probability of a vacuum of level $nm$}

The product of branching ratio's for an upward path of $n$ jumps is

\bea
\prod_{\tilde{n}=1}^{n}\left(\frac{\kappa_{ij}}{D_j}\right)^{up}_{\tilde{n}}&=&
\prod_{\tilde{n}=1}^{n}\left[\frac{\exp
\left[\frac{-24\pi^2}{\Lambda_j}\tilde{r}_j(\Delta
\Lambda_{\tilde{n}})\right]}{\left(\sum_{\ell=1}^{P}\exp\left[\frac{-24\pi^2}{\Lambda_j}
\left[\tilde{r}_j(\Delta\Lambda_{\ell})\right]\right] +
\sum_{q=1}^{\tilde{n}} \exp \left[\frac{-24\pi^2}{\Lambda_j}
\left[\tilde{r}_j(\Delta \Lambda_{q})\right] \right]\right)}\right]
\nonumber
\\&=&\prod_{\tilde{n}=1}^{n}\left[\sum_{\ell=1}^{P}\exp\left[\frac{-24\pi^2}{\Lambda_j}
\left[\tilde{r}_j(\Delta\Lambda_{\ell})-\tilde{r}_j(\Delta
\Lambda_{\tilde{n}})\right]\right] + \sum_{q=1}^{\tilde{n}} \exp
\left[\frac{-24\pi^2}{\Lambda_j} \left[\tilde{r}_j(\Delta
\Lambda_{q})-\tilde{r}_j(\Delta \Lambda_{\tilde{n}})\right]
\right]\right]^{-1} \label{upprod}\eea

where $P$ is the number of plus coordinates in the dominant vacuum
configuration, and we define $\Delta \Lambda_{\tilde{n}}$ to be the
change in $\Lambda$ for the $\tilde{n}'th$ upward jump.  Also, we
have introduced the notation $\tilde{r}_j(\Delta\Lambda_a) \equiv
\tilde{r}(x,y)$ with $y=\frac{2\Lambda_j}{\Delta\Lambda_a}-1$ and
$x=3/4$ (see Eq's. (\ref{ADKy},\ref{newgravfactor})).

The second sum takes into account that for each upward jump, a
coordinate that was $-$ becomes a $+$ coordinate, and therefore
transitions downward in these directions must be included in $D_j$
when calculating the branching ratio.
%(recall when we calculate the branching ratio of an upward
%transition \emph{to} $\Lambda_j$, we calculate the downward
%transition rate \emph{from} $\Lambda_j$, and divide by the sum of
%downward transitions from the target higher-$\Lambda$ vacuum, $D_j$.
%There is also an overall suppression factor which we keep track of
%separately.).

Similarly, the product of branching ratio's for a downward path with
$m$ jumps is

\beqa
\prod_{\tilde{m}=2}^{m}\left(\frac{\kappa_{ij}}{D_j}\right)^{down}_{\tilde{m}}
&=& \prod_{\tilde{m}=2}^{m}\left[\sum_{\ell=1}^{P}\exp
\left[\frac{-24\pi^2}{\Lambda_j} \left[\tilde{r}_j(\Delta
\Lambda_{\ell})-\tilde{r}_j(\Delta\Lambda_{\tilde{m}})\right]\right]
\right. \nonumber \\ &-& \left. \sum_{m'=1}^{\tilde{m}-1} \exp
\left[\frac{-24\pi^2}{\Lambda_j}
\left[\tilde{r}_j(\Delta\Lambda_{m'})-\tilde{r}_j(\Delta
\Lambda_{\tilde{m}})\right] \right] \right.  \nonumber \\ &+& \left.
\sum_{q=1}^{n} \exp \left[\frac{-24\pi^2}{\Lambda_j}
\left[\tilde{r}_j(\Delta \Lambda_{q})-\tilde{r}_j(\Delta
\Lambda_{\tilde{m}})\right] \right]\right]^{-1}
\label{downprod}\eeqa

Since the first downward transition does not have a $D_j$ in the
denominator (see Eq. (\ref{noDj})), our product of branching ratios
starts with the second downward jump, thus $\tilde{m}$ starts at $2$
instead of $1$.

The first sum includes the contribution to $D_j$ from all the
original $P$ $+$ coordinates in $\Lambda_*$.  However, for each jump
down on the return path, the next $\Lambda_j$ will have one less $+$
coordinate and thus the transition rate down in this ``lost''
direction, must not be included in $D_j$.  Thus we subtract the
second sum from the first to account for directions which were
originally $+$'s but in which we can no longer jump down.

The third sum takes into account that the downward path jumps take
place after $n$ upward jumps have changed $n$ coordinates from $-$
to $+$, and therefore transitions downward in these $n$ directions
must be included in $D_j$ when calculating the branching ratio.

Thus the contribution of a path to the prior probability of a vacuum
of level $nm$ is \be P_{nm} \propto
\prod_{\tilde{n}=1}^{n}\left(\frac{\kappa}{D}\right)^{up}
\prod_{\tilde{m}=2}^{m}\left(\frac{\kappa}{D}\right)^{down}
\kappa_{st}\left(\frac{\Lambda_{max}}{\Lambda_*}\right)^2\exp \left[
 -24\pi^2\left(\frac{1}{\Lambda_{*}} - \frac{1}{\Lambda_{max}}\right)\right]
\label{eqn:pn1}
\ee

with $\prod_{\tilde{n}=1}^{n}\left(\frac{\kappa}{D}\right)^{up}$ and
$\prod_{\tilde{m}=2}^{m}\left(\frac{\kappa}{D}\right)^{down}$ as
given by Eq's. (\ref{upprod}) and (\ref{downprod}), and
$\kappa_{st}$ is the first downward jump after having reached
$\Lambda_{max}$ and is given by

\be \label{eqn:kst} \kappa_{st} \approx \exp
\left[\frac{-24\pi^2}{\Lambda_{max}}\tilde{r}_{\Lambda_{max}}(\Delta\Lambda_{\tilde{m}=1})\right]
\ee

Clearly Eq. (\ref{eqn:pn1}) is very complicated.  There are $N_{nm}$
different vacua belonging to level $nm$ and each one has in
principle a unique prior probability which depends on the details of
the transitions made, and not just on the number of up and down
jumps (as in the single step model).  We need to stress the fact
that for the MS model we have a distribution of prior probabilities
for level $nm$ vacua, in contrast to the unique prior probability
$P_n$ for level $n$ vacua in the single-step model.  We cannot
explicitly calculate all the possible probabilities for level $nm$
vacua\footnote{Even for a specific level $nm$ vacuum, there are
$n!m!$ different ways to get to this vacuum, and each path could be
weighted differently.}, so we will consider two limiting cases of
interest to us.

It seems reasonable that if $c$ is ``large enough'', then individual
transition rates are high and do not differ very much.  Thus we
expect that the spread in probabilities for level $nm$ vacua will be
small.  If the spread is small enough we can say that level $nm$
vacua have the same prior probability of $P_{nm}$.  Also different
orderings to reach the same vacuum should become roughly equally
weighted.

On the other hand, if $c$ is ``very small'', then the suppressed
transition rates will differ markedly depending on path details and
there will be a large spread in the prior probabilities of the
different vacua within level $nm$.

In the next subsection we will try to quantify these heuristic
expectations for the large $c$ regime.

The reader may find it useful at this point to turn to Appendix
\ref{AMS} where we study a simpler ``averaged'' version of our
multi-step model analytically.

\subsection{Large c approximation to the prior probability}
\begin{figure}
\begin{center}
\leavevmode\epsfxsize=5in\epsfbox{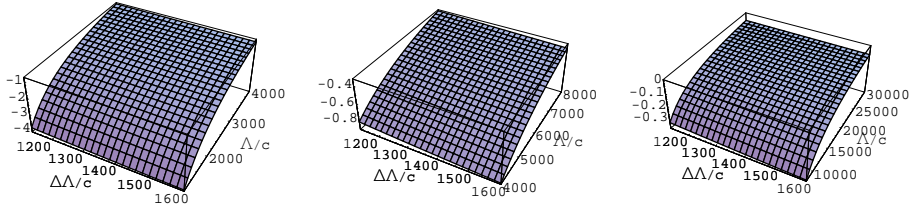}
\end{center}
\caption{The purpose of these plots is to help us estimate the
branching ratios contributing to the \emph{upward} path for
``large'' $c=10^{-2}$. We show three surface plots spanning the
possible $-B_j=\ln{\kappa_{ij}}$  with the following ranges for
$\Lambda$: the first plot has $1133<\Lambda/c<4000$, the second has
$4000<\Lambda/c< 8000$ and the third has $8000<\Lambda/c< 30000$.
All three surfaces have $1133<\Delta\Lambda/c<1600$.  Notice that
along a contour of constant $\Lambda$, there is very little
dependence on $\Delta\Lambda$.  This shows that $\kappa/D \sim$
1/\{number of possible directions in which downward transitions can
be made\}.} \label{largecuppath}
\end{figure}

\begin{figure}
\begin{center}
\leavevmode\epsfxsize=5in\epsfbox{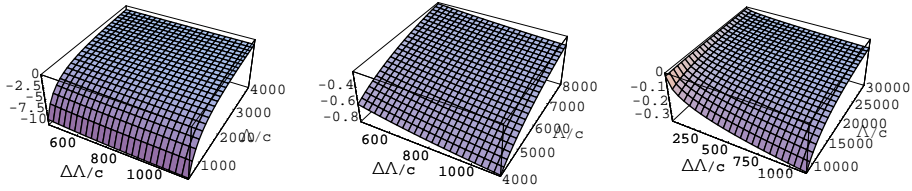}
\end{center}
\caption{The purpose of these plots is to help us estimate the
branching ratios contributing to the \emph{downward} path for
``large'' $c=10^{-2}$. We show three surface plots spanning the
possible $-B_j=\ln{\kappa_{ij}}$ with the following ranges for
$\Lambda/c$ and $\Delta\Lambda/c$ : the first plot has
$450<\Lambda/c<4000$ and $450<\Delta\Lambda/c<1132$, the second has
$4000<\Lambda/c< 8000$ and $450<\Delta\Lambda/c<1132$, while the
third has $8000<\Lambda/c< 30000$ and $1<\Delta\Lambda/c<1132$.
Again we see that along a contour of constant $\Lambda$, the
dependence on $\Delta\Lambda$ is weak.% especially for the
%lower-$\Lambda$ jumps.  For the higher-$\Lambda$ jumps, the constant
%$\Lambda$ curve gets a little steeper, making the overall branching
%ratio typically a little bigger than 1/\{number of possible
%directions in which downward transitions can be made\}. The reason
%why $\kappa/D$ gets bigger is because for high-$\Lambda$ we expect
%the optimal paths to have small steps which have higher transition
%rates than many of the larger jump-size terms contributing to $D_j$.
%This effect is very small for $c=10^{-2}$ and can be seen in Fig.
%\ref{termbytermB}
} \label{largecdownpath}
\end{figure}

For $c \sim 10^{-2}$ all the branching ratios in Eq's.
(\ref{upprod}) and (\ref{downprod}) reduce to 1/\{number of possible
directions in which downward transitions can be made\} (see Fig.'s
\ref{largecuppath} and \ref{largecdownpath}) and the prior
becomes\footnote{The astute reader might notice that in this limit
all the transition rates are only slightly suppressed and the
validity of the semi-classical calculation of the tunneling exponent
may be questionable.  We wish to emphasize that the branching ratios
we have used in calculating the prior should still give the correct
results, even if it is unclear what the exact value of each
individual transition rate is. This can be easily understood as
follows.  If we are in a regime of unsuppressed transitions, then
all allowed transition rates (from a given $\Lambda$) are the same.
Thus the branching ratio for a given transition will just be the
fraction 1/\{number of possible directions in which downward
transitions can be made\}, as we have here.}

\be P_{nm} \propto
n!m!\prod_{\tilde{n}=1}^{n}\left(P+\tilde{n}\right)^{-1}
\prod_{\tilde{m}=2}^{m}\left(P-(\tilde{m}-1)+\tilde{n}\right)^{-1}
\Lambda_{max}^2\exp \left[
 \frac{24\pi^2}{\Lambda_{max}}\left(1-\tilde{r}_{\Lambda_{max}}(\Delta\Lambda_{\tilde{m}=1})\right)\right]
\label{largecpn1} \ee

where we have dropped both factors in Eq. (\ref{eqn:pn1}) which
include $\Lambda_*$ since they will cancel when we take ratios of
probabilities. Also, we have included the factor of $n!m!$ to
account for the different order in which the transitions can be made
to get to each level $nm$ vacuum. Strictly speaking some paths will
be more heavily weighted than others and we should replace the
factor $n!m!$ by some function $f(n,m)$.  We assume here that
$f(n,m)=\beta n!m!$ where we take $\beta$ to be a constant fraction
for any specific value of $c$. Thus $\beta$ will drop out when we
evaluate ratios of probabilities for different values of $nm$.

Noting that

\be \prod_{\tilde{n}=1}^{n}\left(P+\tilde{n}\right)^{-1} =
\frac{P!}{(P+n)!}\ee and \be
\prod_{\tilde{m}=2}^{m}\left(P-(\tilde{m}-1)+n\right)^{-1}=
\frac{(P+n)(P+n-m)!}{(P+n)!}\ee the prior can be written as

\be P_{nm} \propto
n!m!\frac{P!(P+n-m)!(P+n)}{\left((P+n)!\right)^2}\Lambda_{max}^2\exp
\left[
 \frac{24\pi^2}{\Lambda_{max}}\left(1-\tilde{r}_{\Lambda_{max}}(\Delta\Lambda_{\tilde{m}=1})\right)\right]
\label{largecpn2} \ee

\subsection{Small $c$ regime and the prior probability}\label{smallcapproxprior}

Let's look at Fig's. \ref{largecuppath} and \ref{largecdownpath}
again.  If we were to re-plot these surfaces for smaller $c$, the
only thing that would change would be the scale of the height. Since
the tunneling exponent $B_j \propto 1/c$, smaller values of $c$
result in larger $B_j$ and thus more suppression. This indicates
that as $c$ get's smaller there is more of an absolute difference in
the amount of suppression for jumps occuring in different directions
from the same value of $\Lambda$. The net result is that different
paths will have different prior probabilities even when comparing
paths that have the same level $nm$ (this spread in probabilities
for a given level was not present in the single-step and averaged
multi-step models). Thus we cannot come up with a closed form
simplified analytic expression for the prior probability in the
small $c$ case.  But we will show numerically that when $c \sim
10^{-3}$  vacua of the same level, reached via different paths,
start to have prior probabilities which differ by several
e-foldings.  This allows for the possibility of a staggered
distribution.

\subsection{Numerical Interlude}
Using numerics we will reinforce our claim that level $nm$ vacua
have almost identical probabilities even if reached via different
subclasses of trajectories for $c \sim 10^{-2}$. We will also
reinforce our claim that already for $c\sim 10^{-3}$ path dependent
differences cause a spread in probabilities amongst same level $nm$
vacua.  We will consider the vacua reached via $n$ up jumps and
$m=\alpha n$ down jumps with $n = 19, 20$ and $21$. We define three
subclasses of trajectories as follows:
\begin{enumerate}
  \item The ``evenly spread out'' trajectory:
  Let's consider the  $n$ up jumps first.  There are $M=468$ $-$'ve directions to choose from to make
  our $n$ jumps.  Thus we choose a trajectory such that within each interval of consecutive $M/n$
  coordinates there will be one jump.  We assume that each jump
  occurs at the mid-point of the interval.   We calculate the probabilities by ordering the up jumps from largest to smallest.
  We do the same for the down jumps, only now we order jumps from
  smallest to largest and the size of the interval from which we
  pick our coordinates for each jump is given by $P/m$, where $P$ is the number of
  original plus coordinates in the dominant vacuum.
\item The ``extreme'' trajectory:
  Here we jump up in the largest $n/2$ possible directions, and then
  in the smallest $n/2$ directions out of the original $M$
  coordinates. All the up jumps are ordered in decreasing step size.
  For the downward path we order jumps from smallest to largest starting with
  the $m/2$ smallest $+$ coordinates, and then ending with the $m/2$
  largest $+$ directions.
  \item The ``clumped around the average value'' trajectory:
  Again consider the upward path.  We calculate the average value of
  $\Delta\Lambda$ for the $M$ directions we could go up in.  We then
  consider  the $n/2$ coordinates above to the $n/2$ coordinates
  below this average, and assume that our jumps are made from
  largest to smallest out of this consecutive range of integers.
  For the downward path we calculate the average $\Delta\Lambda$ for
  the original $P$ directions we could go down in.   We then
  consider  the $m/2$ coordinates above to the $m/2$ coordinates
  below this average, and assume that our jumps were made from
  smallest to largest out of this consecutive range of integers.

  \end{enumerate}

For each of these trajectories we numerically calculated the prior
probability according to Eq. (\ref{eqn:pn1}) for $c = 10^{-2}$ and
$c=10^{-3}$ for several values of $n$.  The results are displayed in
Tables \ref{tablelargec} and \ref{tablesmallc} where we give the
$-\ln P_{nm}$ for each case. Thus we see that indeed level $nm$
vacua have almost identical probabilities even if reached via
different subclasses of trajectories for $c \sim 10^{-2}$, whilst
for $c \sim 10^{-3}$ there is already a spread in the prior
probabilities amongst level $nm$ vacua.

\begin{table}
\centering
\begin{tabular}{|l|c|c|c|}
 \hline  subclass & n=19 & n=20 & n=21 \\
\hline
  spread & 454.63 & 475.73 & 496.84 \\
  extreme & 453.67 & 474.74 & 495.81 \\
   clumped& 452.57 & 473.68 & 494.78 \\
  \hline
\end{tabular}
 \caption{For $c=10^{-2}$ we find the round trip prior probability and display $-\ln{P_{nm}}$ for the three subclasses
 of trajectories discussed in the text.  Results are shown for various values of $n$ and
$m=\alpha n$. We see that, for a given $nm$, there is essentially no
difference between the different trajectories. But the prior is a
factor of $\sim 10^9$ times greater for $n=19$ than for $n=20$, and
similarly for $n=20$ vs $n=21$. This is easily understood as
follows.  The $n+1$ trajectory has approximately 3 extra jumps over
the $n$ trajectory (one for the extra up direction and roughly 2
extra in the down direction).  Each one of these extra jumps
contributes a factor of $\kappa/D \sim O(10^{-3})$ because there are
roughly $1000$ directions to jump down in.}\label{tablelargec}
\end{table}

\begin{table}
\centering
\begin{tabular}{|l|c|c|c|}
 \hline  subclass & n=19 & n=20 & n=21 \\
\hline
  spread & 545 & 566.58 & 587 \\
  extreme & 528.9 & 549.85 & 570 \\
  clumped & 564.2 & 585.27 & 606 \\
  \hline
\end{tabular}
 \caption{For $c=10^{-3}$ we find the round trip prior probability and display $-\ln{P_{nm}}$ for the three subclasses
 of trajectories discussed in the text.  Results are shown for various values of $n$ and
$m=\alpha n$.  We see a large spread in probabilities amongst same
level $nm$ vacua which makes it impossible for us to use our
procedure for comparing observed probabilities in the anthropic
range as we did in the single-step model, the AMS model discussed in
the appendix, and as we will do in the large c multi-step case in
Section \ref{fulldistributionlargec}.}\label{tablesmallc}
\end{table}

\begin{figure}
\begin{center}
\leavevmode\epsfxsize=5in\epsfbox{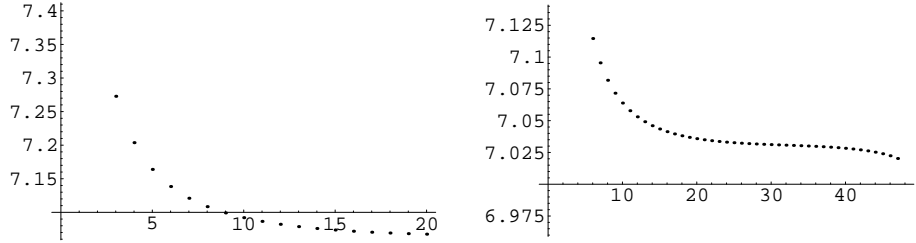}
\end{center}
\caption{ On the left we show  $-\ln(\kappa/D)$ (ie the branching
ratio exponential suppression factors) for the ``spread'' up path
plotted vs the number of the jump.  On the right we show
$-\ln(\kappa/D)$ for the ``spread'' down path plotted vs the
``reverse'' of the number of jumps (for example $10$ means the tenth
to last downward jump and so on). We can see that transitions
occurring from lower-$\Lambda$ (those closer to the y-axis) are more
suppressed. Note also that all these value are pretty similar, which
is to be expected since we are in the large $c$ regime with
$c=10^{-2}$.  The first two up path points are at $\{7.49, 7.30\}$
but are not depicted.  Also, not shown are the four points at
$\{7.44, 7.27, 7.19, 7.14\}$ closest to the y-axis for the downward
path.} \label{termbytermB}
\end{figure}

Let's now take a look at Fig. \ref{termbytermB} where we plot
$-\ln(\kappa/D)$ for each jump in an up and down path with
$(n,m)=(20,48)$ and $c=10^{-2}$.  The jumps up were made in order of
decreasing jump size and the reverse order for the descent back
down.  The first few jumps have the highest values for the
``branching ratio'' suppression (we distinguish between the
suppression caused by the branching ratio factors and the overall
up-jump suppression which depends on how high up in $\Lambda$ we
have to jump before the downward descent) because they are taking
place in lower-$\Lambda$ backgrounds than the later small jumps in
the path \footnote{To be absolutely clear, if we made the necessary
small jumps first they would lead to much greater suppression.  But
any vacuum is reached via a sum of all possible paths which lead to
it, and thus if any large jumps are necessary or possible, then it
is possible to make the large jumps first and this path to a given
vacuum will be larger than paths which entail making the small jumps
first.}. The small jumps all take place where $\Lambda$ is already
relatively large, and their suppression effects are marginalized.
However, we must point out that even if a jump occurs at a high
enough $\Lambda$ such that transition rates are unsuppressed, the
branching ratio for the jump is still a fraction, $\kappa/D \approx
10^{-3}$, so that a path with more steps always pays some penalty.

\begin{figure}
\begin{center}
\leavevmode\epsfxsize=5in\epsfbox{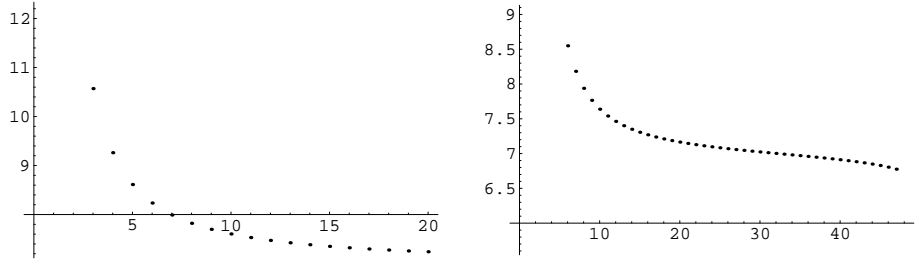}
\end{center}
\caption{ Same as previous plots except now $c=10^{-3}$. Note that
the suppression is larger.  The first few points are not displayed.
For the up path they are $\{26.71, 13.91\}$. The highest few values
for the down path have also been omitted from the display. They are
$\{36.40, 17.65, 12.37, 10.21, 9.14, 8.55,... \}$ for the last jump,
second last jump and so on.} \label{smallctermbytermB}
\end{figure}

%Thus the first many up jumps will be large and approximately the
%same for all the different paths.

\section{Distribution for the observed $\Lambda$ in the large $c$
regime} \label{fulldistributionlargec}

We should now be convinced that in the large $c$ regime vacua of
level $nm$ have the same prior $P_{nm}$.  We thus have all the parts
necessary to calculate the probability of observing each value
$\Lambda_{nm, \ell}$ in a typical realization of our toy model.
These are given by \be P_\obs(\Lambda_{nm, \ell}) \propto P_{nm}
n_\obs(\Lambda_{nm, \ell}) \ee The chance that we live in a world of
a given level $nm$ is then given by \be P_\obs(nm) \propto P_{nm}
\sum_{\ell} n_\obs(\Lambda_{nm, \ell}) \label{eqn:Pobsnm} \ee

We will consider two cases\footnote{Recall from the introduction,
$\Lambda_c \sim 6\times 10^{-120}$ parameterizes the anthropic
suppression.}.

\begin{enumerate}
  \item When $\Lambda_{nm}\ll\Lambda_c$, the sum can be approximated by an
integral, \be \label{eqn:littlesumnm} \sum_{\ell}
n_\obs(\Lambda_{nm, \ell}) \approx \frac{1}{\Delta_{nm}}
\int_0^\infty d\Lambda n_\obs(\Lambda) =
\frac{3\sqrt{\pi}\Lambda_c}{4\Delta_{nm}} =
\frac{3\sqrt{\pi}N_{nm}\Lambda_c}{4\sigma_{nm}} =
\frac{3\sqrt{\pi}\Lambda_c\ln2} {4\Lambda_{nm}} \ee

and we find \beqa \label{eqn:Pobs1nm} P_\obs(nm) &\propto&
\frac{3\sqrt{\pi}\Lambda_c}{4\sigma_{nm}}\frac{M!P!}{(M-n)!(P-m)!}\frac{P!(P+n-m)!(P+n)}{\left((P+n)!\right)^2}\Lambda_{max}^2\nonumber
\\ &\times&\exp \left[
 \frac{24\pi^2}{\Lambda_{max}}\left(1-\tilde{r}_{\Lambda_{max}}(\Delta\Lambda_{\tilde{m}=1})\right)\right]
\eeqa

where we used Eqs.\ (\ref{eqn:Nn}, \ref{eqn:Nupn}, \ref{eqn:Ndownn},
\ref{largecpn2}).  We have not included the terms involving
$\Lambda_*$, which are the same for all $P_\obs(nm)$.  In Eq.\
(\ref{eqn:Pobs1nm}), $P_\obs(nm)$ is a decreasing function of $n$.

  \item When $\Lambda_{nm}>\Lambda_c$, Eq.~(\ref{eqn:Pobsnm})
will be dominated by the first term, and we find \be
\label{eqn:bigsumnm} \sum_{\ell} n_\obs(\Lambda_{nm, \ell}) \approx
e^{-(\Lambda_{nm}/\Lambda_c)^{2/3}} \ee and \be \label{eqn:Pobs2nm}
P_\obs(nm) \propto
\frac{n!m!P!(P+n-m)!(P+n)}{\left((P+n)!\right)^2}\Lambda_{max}^2\exp
\left[
 \frac{24\pi^2}{\Lambda_{max}}\left(1-\tilde{r}_{\Lambda_{max}}(\Delta\Lambda_{\tilde{m}=1})\right)-\left(\frac{\Lambda_{nm}}{\Lambda_c}\right)^{2/3}\right]
\ee
 In Eq.\
(\ref{eqn:Pobs2nm}), $P_\obs(nm)$ increases with increasing $n$
while $n$ is small and the last term in the exponent is dominant,
but it decreases when $n$ is larger and the other terms are
dominant.

\end{enumerate}

\subsection{Comparison of probabilities for level $n+1,\alpha(n+1)+\gamma$
and $n,\alpha n+3+\gamma$ paths} \label{comparison2}

From Table \ref{tableprob} we see that we are comparing
probabilities of vacua which lie adjacent to one another along a
``bold diagonal'' and thus have similar number densities. For
example; for $n=19, \gamma=1$ we would compare probabilities of the
two levels $(n,(\alpha n+3)+\gamma)=(19,\alpha 19+4)$ and
$(n+1,\alpha(n+1)+\gamma)=(20,\alpha 20+1)$. We have introduced the
bookkeeping integer $-5<\gamma<2$.

\begin{enumerate}
  \item When $\Lambda_{n,\alpha n+3+\gamma} > \Lambda_c$ and $\Lambda_{n+1,\alpha(n+1)+\gamma} >
  \Lambda_c$, and using Eq.(\ref{eqn:Pobs2nm})

  \beqa \frac{P_{obs}(n+1,\alpha(n+1)+\gamma)}{P_{obs}(n,\alpha n+3+\gamma)} &\approx& \frac{n+1}{(P+n+1)(P+n)}
  \frac{(\alpha n+\alpha+\gamma)!}{(\alpha
  n+3+\gamma)!}\frac{(P+n+1-\alpha(n+1)-\gamma)!}{(P+n-\alpha n-3-\gamma)!}
  \nonumber \\ &\times& \left(\frac{n+1}{n}\right)^2\exp \left[\left(\frac{\Lambda_{n,\alpha n+3+\gamma}}{\Lambda_c}\right)^{2/3}-\left(\frac{\Lambda_{n+1,\alpha(n+1)+\gamma}}{\Lambda_c}\right)^{2/3}\right]
  \eeqa

  where we have approximated $\left(\frac{\Lambda_{max}(n+1)}{\Lambda_{max}(n)}\right)^2\approx \left(\frac{n+1}{n}\right)^2$.
  For $n \sim 20$, $\Lambda_{n,\alpha n+3+\gamma} \sim 3  \Lambda_{n+1,\alpha(n+1)+\gamma}$, and we find

  \beqa
\frac{P_{obs}(n+1,\alpha(n+1)+\gamma)}{P_{obs}(n,\alpha n+3+\gamma)}
&\approx& \frac{n+1}{(P+n+1)(P+n)}  \frac{(\alpha
n+\alpha+\gamma)!}{(\alpha
n+3+\gamma)!}\frac{(P+n+1-\alpha(n+1)-\gamma)!}{(P+n-\alpha
n-3-\gamma)!} \nonumber
  \\ &\times& \left(\frac{n+1}{n}\right)^2\exp \left[\left(\frac{\Lambda_{n+1,\alpha(n+1)+\gamma}}{\Lambda_c}\right)^{2/3}\right] \nonumber \\ &\approx& 0.1 \exp \left[\left(\frac{\Lambda_{n+1,\alpha(n+1)+\gamma}}{\Lambda_c}\right)^{2/3}\right]
  \eeqa

  thus $P_{obs}(n+1,\alpha(n+1)+\gamma)\gg P_{obs}(n,\alpha
  n+3+\gamma)$.  Therefore, in this regime $P_{obs}$ grows with decreasing
  $\Lambda$.

  \item When $\Lambda_{n+1,\alpha(n+1)+\gamma} < \Lambda_c$ and $\Lambda_{n,\alpha n+3+\gamma} < \Lambda_c$,  we
  find using Eq. (\ref{eqn:Pobs1nm})
  \beqa \frac{P_{obs}(n+1,\alpha(n+1)+\gamma)}{P_{obs}(n,\alpha
n+3+\gamma)} &\approx& \frac{n+1}{(P+n+1)(P+n)}  \frac{(\alpha
n+\alpha+\gamma)!}{(\alpha n+3+\gamma)!}\nonumber \\ &\times&
\left(\frac{n+1}{n}\right)^2\frac{(P+n+1-\alpha(n+1)-\gamma)!}{(P+n-\alpha
n-3-\gamma)!}\frac{\Lambda_{n,\alpha n+3+\gamma}}{\Lambda_{n+1,\alpha(n+1)+\gamma}} \nonumber \\
&\approx& 0.3\eeqa

\end{enumerate}

Thus we find that several values of $nm$ contribute nearly equally
to the total probability. The first of these might be dominated by a
single $\Lambda_{n, \alpha n +3+\gamma}$, but there will be other
vacua with closer spaced $\Lambda$. These vacua have similar
$n_\obs$ and identical prior probability, so we could easily be in
any of them.

Thus when $c$ is large, we recover approximately the original
anthropic predictions with a smooth prior $P(\Lambda)$.  There might
be an effect due to the discrete nature of the vacua associated with
the smallest $(n, \alpha n +3+\gamma)$, where $P_\obs(n, \alpha n
+3+\gamma)$ has its peak, but this effect is small because level
$(n, \alpha n +3+\gamma)$ does not dominate the probability
distribution. Instead the probability is divided across many
different levels, while only level $n, \alpha n +3+\gamma$ has the
above effect. In other words, we may be more likely to be in the
first $(n,\alpha n+3+\gamma)$ vacuum, but a good fraction of the
probability is still spread amongst many other vacua.  Thus the
distribution is effectively flat.

\subsection{Comparison of probabilities for level $nm$ and $n,(m+1)$
paths} \label{comparison}

From Table \ref{tableprob} we see that level $n,(m+1)$ lie one row
above the level $nm$ vacua.  These vacua have number densities
$\rho_{nm}$ which differ by roughly a factor of $20$.  We will now
evaluate the ratio of probabilities for vacua with levels which
differ via one extra jump down.

\begin{enumerate}
  \item When $\Lambda_{nm} > \Lambda_c$ and $\Lambda_{n(m+1)} >
  \Lambda_c$, and using Eq. (\ref{eqn:Pobs2nm})
  \be \frac{P_{obs}(n(m+1))}{P_{obs}(nm)} =\frac{m+1}{P+n-m}
  \exp \left[\left(\frac{\Lambda_{nm}}{\Lambda_c}\right)^{2/3}-\left(\frac{\Lambda_{n(m+1)}}{\Lambda_c}\right)^{2/3}\right]   \ee
  For  $n \sim 20, m \sim 50$, $\Lambda_{nm} \sim 20
  \Lambda_{n(m+1)}$,
  thus $P_{obs}(n(m+1))\gg P_{obs}(nm)$

  \item When $\Lambda_{nm}>\Lambda_c$ and $\Lambda_{n(m+1)}<\Lambda_c$,  using Eq.'s (\ref{eqn:Pobs1nm},\ref{eqn:Pobs2nm}) we find \be
 \frac{P_{obs}(n(m+1))}{P_{obs}(nm)} =\frac{m+1}{P+n-m}\frac{3 \sqrt{\pi}\Lambda_c\ln{2}}{4\Lambda_{n(m+1)}}
  \exp \left[\left(\frac{\Lambda_{nm}}{\Lambda_c}\right)^{2/3}\right] \approx \frac{1}{20}  \exp \left[\left(\frac{\Lambda_{nm}}{\Lambda_c}\right)^{2/3}\right] \ee
  and therefore $P_{obs}(n(m+1))\gtrsim P_{obs}(nm)$.
  \item When  $\Lambda_{n(m+1)} < \Lambda_c$ and $\Lambda_{nm} <
  \Lambda_c$, using Eq. (\ref{eqn:Pobs1nm}), we find
  \be \frac{P_{obs}(n(m+1))}{P_{obs}(nm)} =\frac{m+1}{P+n-m}\frac{\Lambda_{nm}}{\Lambda_{n(m+1)}}\approx 1.\ee

\end{enumerate}

We can also have vacua reached via $m+2$, $m+3$, ... jumps down and
all these vacua contribute nearly equally to the total probability.
We could live in any one of these.  Thus since level $nm$ doesn't
dominate over level $n(m+1)$, $n(m+2)$,... we again see evidence for
an effectively flat distribution.

%\section{Distribution for the observed $\Lambda$ in the small c
%regime} \label{fulldistributionsmallc}

%\section{Qualitative comparison of single-step and multi-step prior distributions}\label{singlemulticompare}

\section{Discussion}\label{discussion}

A key ingredient in the anthropic prediction of the cosmological
constant is the assumption of a flat prior distribution.  However,
the first attempt to calculate this distribution for the
Bousso-Polchinski and Arkani-Hamed-Dimopolous-Kachru landscape
models \cite{SPV,SP} revealed a staggered distribution, suggesting a
conflict with anthropic predictions.

These calculations were constrained by computational limitations and
revealed only the probabilities of a handful of the most probable
vacua\footnote{It is extremely unlikely that any of these vacua
should lie in the anthropic range.}. In Ref.\ \cite{OSP} we went
beyond these first order perturbative results by studying a simple
toy model which allowed analytic calculation with a large, realistic
number of vacua, $N \sim 10^{500}$. We found an interesting fractal
distribution for the prior $P(\Lambda)$. When including anthropic
selection effects to determine $P_{obs}(\Lambda)$, we found that
agreement with observation depends on the only free parameter of the
model, the jump size $\tilde{c}$.

We showed that when $\tilde{c} \sim 1$, anthropic reasoning does
indeed solve the cosmological constant problem.  Even though the
prior distribution has a rich fractal structure, the states of
interest have similar vacua sufficiently closely spaced to
approximate the flat distribution well enough to give the usual
anthropic results.

On the other hand, if $\tilde{c}$ is small, of order $10^{-3}$, then
the agreement with observation was found to break down. In this case
the volume fraction of universes with large cosmological constant is
so high that in the overall probability they are greatly preferred
even though the selection factor disfavors them.

We emphasize that the primary cause of the massive spread in
probabilities in \cite{SPV,SP} which has been referred to as
staggering, comes from differences in total up-jump suppression
factors. This effect was studied analytically in \cite{OSP}. Jumping
to a higher $\Lambda_{max}$ causes any of the descendant vacua to be
suppressed relative to descendants of lower $\Lambda_{max}$  paths.
Since $\Lambda_{max}$ depends on the free parameter $\tilde{c}$, so
does the relative suppression, and thus we can always tune
$\tilde{c}$ for smooth or staggered distributions.

In this paper we have studied a multi-step model with a \emph{range}
of jump sizes,$|\Delta\Lambda_i |= c i $ and $1 \leq i \leq J$,
parameterized by an overall free parameter $c$. (We also studied a
simpler averaged version of the model, which allowed more progress
analytically, especially for the small $c$ scenario.) In \cite{OSP}
we conjectured that, for such a model, landscapes with large jumps
are more likely to give the standard anthropic results, while those
with small jumps are likely to predict universes unlike ours.  We
argued that the transition rates between vacua all have terms
proportional to $1/c$ in the exponent. Thus when $c$ is small, the
transition rates and the probabilities of different vacua are very
sensitive to the details of the transitions. When $c$ is large, all
rates are larger and less variable.  This conjecture seems to be
borne out by our study of the multi-step and AMS models here.

The conclusion for the multi-jump model, that we can find cases
where anthropic reasoning is both valid and invalid depending on
some free parameter $c$ is analogous to the conclusion reached for
the "single-jump" model \cite{OSP}. However, there are interesting
qualitative differences in the behavior of the prior distributions.
Perhaps we could be so bold as to turn the anthropic argument
around, and use these results to predict what a sensible value for
$c$ should be?

We should say a few words about boundary conditions at the regular
Planck regime, $\Lambda \sim \mathcal{O}(M_{p}^4)$. In particular,
it may be that for the underlying fundamental theory, which we
assume gives rise to the effective potential we have used in our ADK
model, fluxes decompactify at the Planck scale and the effective
potential asymptotes off to zero. If the fluxes do indeed
decompactify it would seem like this Planck boundary would act as a
sink for probability.  We have not included these effects in our
calculation because all of our transitions have taken place within
the Planck boundary since even for $c=10^{-2}$ and $n\sim 20$,
$\Lambda_{max}< 320 M_{P}^4 \sim 0.5 G^{-2}.$

One more speculative thought.  If you consider a multi-jump model
with less of a difference between the ratio of the largest to
smallest jump in the range (in the current model the ratio is 1600
which is pretty big )  this could lead to a smoothing of the prior
distribution for a reasonably small value of $c$.  If the ratio of
maximum and minimum jump sizes were constrained to be say within an
order of magnitude, it could turn out that we can define a unique
prior for a given level $nm$ vacuum, in the small $c$ case too.  In
so doing we would then be able to show that the observed
distribution smooths out by comparing $P_{obs}$ for levels $nm$ and
$n,m+1$ as we did in the large $c$ case.

\appendix

%\section{More on the dominant vacuum configuration}\label{Dominantvacuum}

%This is because there is no special reason to think that the
%dominant vacuum $D_*$ should not have some mix of $+$ and $-$
%coordinates for the small jumps\footnote{Recall for the larger jumps
%we argued that the dominant vacuum configuration should have mostly
%$-$'s, because this would mean that the least suppressed transitions
%are not possible (\textbf{see...}).}.

\section{The ``averaged'' multi-step model}\label{AMS}

The dominant vacuum of the multi-step model we have been considering
has roughly the following structure:

\be \Lambda_*=c\{+.....+-...-\}\ee with $P=0.71 J \approx 1132$
$+$'s and $M=0.29 J\approx 468$ $-$'s.  Jumps down from the plus
coordinates range in size from $1c \leq \Delta \Lambda \leq 1132c$
and jumps up from the minus coordinates range in size from $1133c
\leq \Delta \Lambda \leq 1600c$.

The average jump size of the plus and minus coordinates respectively
are $\Delta\bar{\Lambda}_P \approx 566 c$ and $\Delta\bar{\Lambda}_M
\approx 1366 c$. We will define an ``averaged'' multi-step model
(AMS model) which has a dominant vacuum with $P$ plus coordinates
all having the same jump size $\Delta\bar{\Lambda}_P$ and $M$ minus
coordinates all having the same jump size $\Delta\bar{\Lambda}_M$.

One of the simplifying characteristics of the AMS model is that
vacua reached via $n$ up jumps and $m$ down jumps from the dominant
vacuum all have the same prior probability, $P_n$, which we will now
calculate.

\subsection{The prior probability of the AMS model}

The product of branching ratio's for an upward path of $n$ jumps is

\be
\prod_{\tilde{n}=1}^{n}\frac{\kappa_{ij}}{D_j}=\prod_{\tilde{n}=1}^{n}\left[P\exp\left[\frac{-24\pi^2}{\Lambda_j}
\left[\tilde{r}_j(\Delta\bar{\Lambda}_P)-\tilde{r}_j(\Delta\bar{\Lambda}_M)\right]\right]+\tilde{n}
\right]^{-1} \label{avupprod}\ee

where $\Lambda_j=\tilde{n}\Delta\bar{\Lambda}_M+\Lambda_*$.

The product of branching ratio's for the downward path with $m$
jumps is

\be \prod_{\tilde{m}=2}^{m}\frac{\kappa_{ij}}{D_j} =
\prod_{\tilde{m}=2}^{m}\left[\left(P-(\tilde{m}-1)\right) + n \exp
\left[\frac{-24\pi^2}{\Lambda_j}
\left[\tilde{r}_j(\Delta\bar{\Lambda}_M)-\tilde{r}_j(\Delta\bar{\Lambda}_P)\right]
\right]\right]^{-1} \label{avdownprod}\ee

and in this expression $\Lambda_j =
n\Delta\bar{\Lambda}_M+\Lambda_*-(\tilde{m}-1)\Delta\bar{\Lambda}_P
= \Lambda_{max}-(\tilde{m}-1)\Delta\bar{\Lambda}_P$.

Thus the prior probability of a vacuum of level $n$ is \be P_n
\propto
n!m!\prod_{\tilde{n}=1}^{n}\left(\frac{\kappa}{D}\right)^{up}
\prod_{\tilde{m}=2}^{m}\left(\frac{\kappa}{D}\right)^{down}
\kappa_{st}\left(\frac{\Lambda_{max}}{\Lambda_{*}}\right)^2\exp
\left[
 -24\pi^2\left(\frac{1}{\Lambda_{*}} - \frac{1}{\Lambda_{max}}\right)\right]
\label{eqn:avpn1} \ee

with $\prod_{\tilde{n}=1}^{n}\left(\frac{\kappa}{D}\right)^{up}$ and
$\prod_{\tilde{m}=2}^{m}\left(\frac{\kappa}{D}\right)^{down}$ as
given by Eq's. (\ref{avupprod}) and (\ref{avdownprod}), and
$\kappa_{st}$ is the first downward jump after having reached
$\Lambda_{max}$ and is given by

\be \label{eqn:avkst} \kappa_{st} \approx \exp
\left[\frac{-24\pi^2}{\Lambda_{max}}\tilde{r}_{\Lambda_{max}}(\Delta\bar{\Lambda}_{P})\right]
\ee

Notice that we have included an extra factor of $n!m!$ to account
for all the equally weighted paths that lead to the same vacuum
because the up jumps and down jumps can be taken in any order.

\subsection{Large c approximation to the prior probability of the AMS model}
For $c \sim 10^{-2}$ all the exponential factors in Eq's.
(\ref{avupprod}) and (\ref{avdownprod}) reduce to $1$ and the prior
becomes

\be P_n \propto
n!m!\prod_{\tilde{n}=1}^{n}\left(P+\tilde{n}\right)^{-1}
\prod_{\tilde{m}=2}^{m}\left(P-(\tilde{m}-1)+n\right)^{-1}
\Lambda_{max}^2\exp \left[
 \frac{24\pi^2}{\Lambda_{max}}\left(1-\tilde{r}_{\Lambda_{max}}(\Delta\bar{\Lambda}_{P})\right)\right]
\label{largecavpn1} \ee

where we have dropped the $\Lambda_*$ terms appearing in Eq.
(\ref{eqn:avpn1}).

Noting that

\be \prod_{\tilde{n}=1}^{n}\left(P+\tilde{n}\right)^{-1} =
\frac{P!}{(P+n)!}\ee and \be
\prod_{\tilde{m}=2}^{m}\left(P-(\tilde{m}-1)+n\right)^{-1}=
\frac{(P+n)(P+n-m)!}{(P+n)!}\ee the prior can be written as

\be P_n \propto
n!m!\frac{P!(P+n-m)!(P+n)}{\left((P+n)!\right)^2}\Lambda_{max}^2\exp
\left[
 \frac{24\pi^2}{\Lambda_{max}}\left(1-\tilde{r}_{\Lambda_{max}}(\Delta\bar{\Lambda}_{P})\right)\right]
\label{largecavpn2} \ee

\subsection{Small c approximation to the prior probability of the AMS model}
For $c \sim 10^{-4}$ the first exponential factor in Eq.
(\ref{avupprod}) becomes large and we can ignore the second term. In
Eq. (\ref{avdownprod}) the second term becomes very small and can be
ignored so the prior becomes

\beqa P_n &\propto&
n!m!\prod_{\tilde{n}=1}^{n}\left[P\exp\left[\frac{-24\pi^2}{\Lambda_j}
\left[\tilde{r}_j(\Delta\bar{\Lambda}_P)-\tilde{r}_j(\Delta\bar{\Lambda}_M)\right]\right]
\right]^{-1}
\prod_{\tilde{m}=2}^{m}\left(P-(\tilde{m}-1)\right)^{-1} \nonumber \\
&& \Lambda_{max}^2\exp \left[
 \frac{24\pi^2}{\Lambda_{max}}\left(1-\tilde{r}_{\Lambda_{max}}(\Delta\bar{\Lambda}_{P})\right)\right]
\label{smallcavpn1} \eeqa

where $\Lambda_j=\tilde{n}\Delta\bar{\Lambda}_M+\Lambda_*$ above.

Noting that

\be
\prod_{\tilde{m}=2}^{m}\left(P-(\tilde{m}-1)\right)^{-1}=\frac{P(P-m)!}{P!}\ee
we finally have

\be P_n \propto
n!m!\frac{P(P-m)!}{P^nP!}\prod_{\tilde{n}=1}^{n}\left[\exp\left[\frac{-24\pi^2}{\Lambda_j}
\left[\tilde{r}_j(\Delta\bar{\Lambda}_M)-\tilde{r}_j(\Delta\bar{\Lambda}_P)\right]\right]
\right] \Lambda_{max}^2\exp \left[
 \frac{24\pi^2}{\Lambda_{max}}\left(1-\tilde{r}_{\Lambda_{max}}(\Delta\bar{\Lambda}_{P})\right)\right]
\label{smallcavpn1} \ee

\section{Distribution for the observed $\Lambda$ in the AMS model}
\label{AMSfulldistribution}

We now have all the parts necessary to calculate the probability of
observing each value $\Lambda_{n, \ell}$ in a typical realization of
our toy AMS model. These are given by \be P_\obs(\Lambda_{n, \ell})
\propto P_n n_\obs(\Lambda_{n, \ell}) \ee The chance that we live in
a world of a given level $n$ is then given by \be P_\obs(n) \propto
P_n \sum_{\ell} n_\obs(\Lambda_{n, \ell}) \label{eqn:Pobsn} \ee

We will consider two cases.

\begin{enumerate}
  \item When $\Lambda_n\ll\Lambda_c$, the sum can be approximated by an
integral, \be \label{eqn:littlesum} \sum_{\ell} n_\obs(\Lambda_{n,
\ell}) \approx \frac{1}{\Delta_n} \int_0^\infty d\Lambda
n_\obs(\Lambda) = \frac{3\sqrt{\pi}\Lambda_c}{4\Delta_n} =
\frac{3\sqrt{\pi}N_n\Lambda_c}{4c} = \frac{3\sqrt{\pi}\Lambda_c\ln2}
{4\Lambda_n} \ee

where we have used $\sigma_{nm} \equiv c$ and therefore
$N_n/c=1/\Delta_n$.  We dropped the extra subscript of $m$ for
$N_{nm}$ and $\Delta_{nm}$ since in the AMS model, once we specify
$n$, $m=\alpha n$ is specified too.

  \item When $\Lambda_n>\Lambda_c$, Eq.~(\ref{eqn:Pobsn})
will be dominated by the first term, and we find \be
\label{eqn:bigsum} \sum_{\ell} n_\obs(\Lambda_{n, \ell}) \approx
e^{-(\Lambda_n/\Lambda_c)^{2/3}} \ee
\end{enumerate}

\subsection{Large c regime}

\begin{enumerate}
  \item For $\Lambda_n\ll\Lambda_c$, and including Eqs.\ (\ref{eqn:Nn}, \ref{eqn:Nupn}, \ref{eqn:Ndownn},
\ref{largecavpn2}), we find \be \label{eqn:Pobs1} P_\obs(n) \propto
\frac{3\sqrt{\pi}\Lambda_c}{4c}\frac{M!P!}{(M-n)!(P-m)!}\frac{P!(P+n-m)!(P+n)}{\left((P+n)!\right)^2}\Lambda_{max}^2\exp
\left[
 \frac{24\pi^2}{\Lambda_{max}}\left(1-\tilde{r}_{\Lambda_{max}}(\Delta\bar{\Lambda}_{P})\right)\right]
\ee

We have not included the terms involving $\Lambda_*$, which are the
same for all $P_\obs(n)$.  In Eq.\ (\ref{eqn:Pobs1}), $P_\obs(n)$ is
a decreasing function of $n$.

  \item When $\Lambda_n>\Lambda_c$,  using Eq.\
(\ref{largecavpn2}), we find \be \label{eqn:Pobs2} P_\obs(n) \propto
\frac{n!m!P!(P+n-m)!(P+n)}{\left((P+n)!\right)^2}\Lambda_{max}^2\exp
\left[
 \frac{24\pi^2}{\Lambda_{max}}\left(1-\tilde{r}_{\Lambda_{max}}(\Delta\bar{\Lambda}_{P})\right)-\left(\frac{\Lambda_n}{\Lambda_c}\right)^{2/3}\right]
\ee
 In Eq.\
(\ref{eqn:Pobs2}), $P_\obs(n)$ increases with increasing $n$ while
$n$ is small and the last term in the exponent is dominant, but it
decreases when $n$ is larger and the other terms are dominant.
\end{enumerate}

The division between regimes occurs when $\Lambda_n\sim\Lambda_c$,
i.e., \be c\ln2/N_n \sim \Lambda_c\sim 6\times10^{-120} \ee With
$c\sim 10^{-2}$, we find $n\sim 19/20$.  The dependence on $c$ is
weak, with $c\sim 10^{-4}$ corresponding to $n\sim 19$.  For $n$ in
this range, changing $n$ by one unit changes $\Lambda_n$ by a factor
of about $10^4$.  Thus there is at most one $n$ with
$\Lambda_n\sim\Lambda_c$.

Comparing Eq.'s\ (\ref{eqn:Pobs1}) and Eq.\ (\ref{eqn:Pobs2}) for
the same $n$, we see that they differ by a factor of $\Lambda_cN_n/c
\exp{(\Lambda_n/\Lambda_c)^{2/3}}\sim 1$ if $\Lambda_n \sim
\Lambda_c$, so there is no big jump due to switching regimes.

Now let us start with $n = 1$ and increase $n$.  Certainly with $n =
1$, $\Lambda_n\gg\Lambda_c$ by a huge factor, we are in the regime
of Eq.\ (\ref{eqn:Pobs2}), and $P_\obs$ is infinitesimal.  As we
increase $n$, $P_\obs$ increases.  Once $n$ is significantly above
1, we can approximate the increase from one step to the next as
\beqa \label{eqn:pp1} \frac{P_\obs(n+1)}{P_\obs(n)}&\approx&
\left(\frac{(n+1)(\alpha(n+1))!(P+n+1-\alpha(n+1))!}{(P+n+1)(\alpha
n)!(P+n-\alpha n)!(P+n)}\right) \nonumber \\&\times&
\left(\frac{n+1}{n}\right)^2\exp\left[-\frac{24\pi^2}{cn^2\Delta\bar{\Lambda}_M}(1-\tilde{r}(\Delta\bar{\Lambda}_P))
+\left(\frac{\Lambda_n}{\Lambda_c}\right)^{2/3}\right] \eeqa where
we have ignored $\left(\Lambda_{n+1}/\Lambda_c\right)^{2/3}$ as much
less than $\left(\Lambda_n/\Lambda_c\right)^{2/3}$ and used
\be\left(\frac{\Lambda_{max}(n+1)}{\Lambda_{max}(n)}\right)^2=\left(\frac{n+1}{n}\right)^2\ee
$\sim 1.1$ for parameters of interest. In the first term
$\tilde{r}(\Delta\bar{\Lambda}_P) << 1$ for $n\sim 19$, so we will
ignore this contribution.

Also, since we are in the regime of large $c \approx 10^{-2}$, for
$J = 1600$, $n\sim 19$, the first term in the exponent is $<-1$.
Thus, for sufficiently small $n$, the last term in the exponent
dominates and $P_\obs(n+1)/P_\obs(n)\gg 1$.  There is only an
infinitesimal probability that we will be in a vacuum of level $n$,
because there are others that are much more probable.  As we
increase $n$, $P_\obs(n)$ will continue to increase.

By the time we reach the point where $\Lambda_n\sim\Lambda_c$ the
exponential term becomes insignificant.

In that case, we switch to the regime of Eq.\ (\ref{eqn:Pobs1}),
where $P_\obs$ decreases only slowly with increasing $n$.  In this
regime, \be \frac{P_\obs(n+1)}{P_\obs(n)}\approx \frac{N_{n+1}}{N_n}
\frac{P_{n+1}}{P_n} \left(\frac{n+1}{n}\right)^2\ee which is about
$0.41$ for parameters of interest.  Thus we find that several values
of $n$ contribute nearly equally to the total probability. The first
of these might be dominated by a single $\Lambda_n$, but the others
will have a large number of closely spaced $\Lambda$. These vacua
have similar $n_\obs$ and identical prior probability, so we could
easily be in any of them.

Thus when $c$ is large, we recover approximately the original
anthropic predictions with a smooth prior $P(\Lambda)$.  There might
be an effect due to the discrete nature of the vacua associated with
the smallest $n$, where $P_\obs(n)$ has its peak, but this effect is
small because level $n$ does not dominate the probability
distribution.  Instead the probability is divided across many
different levels, while only level $n$ has the above effect.

\subsection{Small c regime}

\begin{enumerate}
  \item For $\Lambda_n\ll\Lambda_c$, and including Eqs.\ (\ref{eqn:Nn}, \ref{eqn:Nupn}, \ref{eqn:Ndownn},
\ref{smallcavpn1}), we find \beqa \label{eqn:smallPobs1} P_\obs(n)
&\propto&
\frac{3\sqrt{\pi}\Lambda_c}{4c}\frac{M!}{(M-n)!}\frac{P}{P^n}\Lambda_{max}^2\exp
\left[
 \frac{24\pi^2}{\Lambda_{max}}\left(1-\tilde{r}_{\Lambda_{max}}(\Delta\bar{\Lambda}_{P})\right)\right]\nonumber \\ &\times&\prod_{\tilde{n}=1}^{n}\left[\exp\left[\frac{-24\pi^2}{\Lambda_j}
\left[\tilde{r}_j(\Delta\bar{\Lambda}_M)-\tilde{r}_j(\Delta\bar{\Lambda}_P)\right]\right]
\right]\eeqa

In Eq.\ (\ref{eqn:smallPobs1}), $P_\obs(n)$ is a decreasing function
of $n$.

  \item When $\Lambda_n>\Lambda_c$,  using Eq.\
(\ref{smallcavpn1}), we find \beqa \label{eqn:smallPobs2} P_\obs(n)
&\propto& \frac{n!m!P(P-m)!}{P^n P!}\Lambda_{max}^2\exp \left[
 \frac{24\pi^2}{\Lambda_{max}}\left(1-\tilde{r}_{\Lambda_{max}}(\Delta\bar{\Lambda}_{P})\right)-\left(\frac{\Lambda_n}{\Lambda_c}\right)^{2/3}\right] \nonumber \\ &\times& \prod_{\tilde{n}=1}^{n}\left[\exp\left[\frac{-24\pi^2}{\Lambda_j}
\left[\tilde{r}_j(\Delta\bar{\Lambda}_M)-\tilde{r}_j(\Delta\bar{\Lambda}_P)\right]\right]
\right] \eeqa
 In Eq.\
(\ref{eqn:smallPobs2}), $P_\obs(n)$ increases with increasing $n$
while $n$ is small and the last term in the exponent is dominant,
but it decreases when $n$ is larger and the other terms are
dominant.
\end{enumerate}

Now let us start with $n = 1$ and increase $n$.  Again, with $n =
1$, $\Lambda_n\gg\Lambda_c$ by a huge factor, we are in the regime
of Eq.\ (\ref{eqn:smallPobs2}), and $P_\obs$ is infinitesimal.  As
we increase $n$, $P_\obs$ increases.  Once $n$ is significantly
above 1, we can approximate the increase from one step to the next
as \be \label{eqnsmall:pp1} \frac{P_\obs(n+1)}{P_\obs(n)}\approx
\frac{n+1}{P}\frac{(\alpha(n+1))!}{(\alpha
n)!}\frac{(P-\alpha(n+1))!}{(P-\alpha n)!}
\left(\frac{n+1}{n}\right)^2\exp\left[-\frac{24\pi^2}{n^2\Delta\bar{\Lambda}_M}
+\left(\frac{\Lambda_n}{\Lambda_c}\right)^{2/3}\right] \ee where we
have ignored $\left(\Lambda_{n+1}/\Lambda_c\right)^{2/3}$ as much
less than $\left(\Lambda_n/\Lambda_c\right)^{2/3}$. (For $1<n<21$
the prefactor is $\mathcal{O}(10^{-2}/10^{-3})$).

For sufficiently small $n$, the last term in the exponent dominates
and $P_\obs(n+1)/P_\obs(n)\gg 1$.  There is only an infinitesimal
probability that we will be in a vacuum of level $n$, because there
are others that are much more probable.  As we increase $n$,
$P_\obs(n)$ will continue to increase. What happens next depends on
the magnitude of $c$.

Let's suppose that \be \label{eqn:csmall} c <
\left(\frac{\Lambda_{n+1}}{\Lambda_n}\right)^{2/3}\frac{24\pi^2}{n^{2}\left(\Delta\bar{\Lambda}_M/c\right)}
\ee for relevant values of $n$.

Thus we can find a value of $n$ such that \be \label{eqn:ll} 1 <
\left(\frac{\Lambda_{n+1}}{\Lambda_n}\right)^{2/3}\frac{24\pi^2}{n^{2}\left(\Delta\bar{\Lambda}_M\right)}<
\left(\frac{\Lambda_n}{\Lambda_c}\right)^{2/3}
<\frac{24\pi^2}{n^{2}\left(\Delta\bar{\Lambda}_M\right)} \ee We will
now show that for this $n$, \be \label{eqn:PvsP} P_\obs(n+1)\ll
P_\obs(n)\,, \ee so that we should find ourselves in a vacuum of at
most level $n$.

It is not clear from Eq.\ (\ref{eqn:ll}) whether
$\Lambda_{n+1}/\Lambda_c$ is more or less than 1, so we might need
to use either Eq.\ (\ref{eqn:smallPobs1}) or Eq.\
(\ref{eqn:smallPobs2}) for $P_\obs(n+1)$.  We will prove the claim
using Eq.\ (\ref{eqn:smallPobs1}). Since this gives a larger value
than Eq.\ (\ref{eqn:smallPobs2}), if Eq.\ (\ref{eqn:PvsP}) holds
using Eq.\ (\ref{eqn:smallPobs1}), it will certainly hold using Eq.\
(\ref{eqn:smallPobs2}).  Thus we will take \be \label{eqn:pp2}
\frac{P_\obs(n+1)}{P_\obs(n)} \approx \frac{3\sqrt{\pi}\Lambda_c
\ln{2}}{4\Lambda_{n+1}} \frac{(n+1)}{P}\frac{(\alpha(n+1))!}{(\alpha
n)!}\frac{(P-\alpha(n+1))!}{(P-\alpha n)!}
\left(\frac{n+1}{n}\right)^2\exp\left[-\frac{24\pi^2}{n^2\Delta\bar{\Lambda}_M}
+\left(\frac{\Lambda_n}{\Lambda_c}\right)^{2/3}\right] \ee

Now from Eq.\ (\ref{eqn:csmall}), we find that \be
\frac{24\pi^2}{n^2\Delta\bar{\Lambda}_M}>
\left(\frac{\Lambda_n}{\Lambda_{n+1}}\right)^{2/3}\approx
\left(\frac{N_{n+1}}{N_{n}}\right)^{2/3} \approx 10^{8/3} \ee Thus
unless $(\Lambda_n/\Lambda_c)^{2/3}$ is extremely close to the upper
bound in Eq.\ (\ref{eqn:ll}), the exponential term in Eq.\
(\ref{eqn:pp2}) will be infinitesimal, and Eq.\ (\ref{eqn:PvsP})
will follow.  If we do have $(\Lambda_n/\Lambda_c)^{2/3}\approx
24\pi^2/(n^2\Delta\bar{\Lambda}_M)$, then
$\Lambda_{n+1}/\Lambda_c\agt 1$.  Then the prefactors in Eq.\
(\ref{eqn:pp2}) are at most about $10^{-2}$ for parameters of
interest, and again Eq.\ (\ref{eqn:PvsP}) follows.

Thus we can say with great confidence that we live in a universe
with level $n$ or lower.  From Eq.\ (\ref{eqn:ll}), we see
immediately that we should observe $\Lambda\ge\Lambda_c$, whereas in
fact we observe $\Lambda_0\approx 0.1\Lambda_c$.  If $c$ is
significantly smaller than the limit in Eq.\ (\ref{eqn:csmall}),
then we will see a very ``non-anthropic'' universe.

%We will be able to find $\Lambda_n$ with
%$(\Lambda_n/\Lambda_c)^{2/3}> 24\pi^2/(cn^{2/3}J^{4/3})$, and thus
%\be n_\obs(\Lambda) \alt n_\obs(\Lambda_n) <
%e^{-24\pi^2/(cn^{2/3}J^{4/3})} \ee will be tiny, meaning that only
%an infinitesimal fraction of matter has coalesced into galaxies.
%For example, with $c=10^{-3}$, we would find $n_\obs(\Lambda)\alt
%e^{-3}\approx 0.05$, in contrast to the observed value (in our
%approximation) $n_\obs(\Lambda)\approx 0.85$. With $c=10^{-4}$, we
%would find $n_\obs(\Lambda)\alt e^{-30}\approx 10^{-13}$: a universe
%utterly unlike our own.

\section{Bubble abundances} \label{abundances}

We review here the procedure for calculating the volume fraction of
vacua of a given kind, using the ``pocket-based measure'' formalism
of Refs.\ \cite{GSPVW,SPV}.

Transition rates depend on the details of the landscape. However, if
$\Lambda_i < \Lambda_j$, the rate of the transition upward from $i$
to $j$ is suppressed relative to the inverse, downward transition,
by a factor which does not depend on the details of the process
\cite{EWeinberg}\footnote{We assume only Lee-Weinberg tunnelings and
do not consider Farhi-Guth-Guven (FGG) tunnelings \cite{FGG}. These
FGG tunnelings may be faster in upward transition rates, but their
interpretation is unclear \cite{AJ,AGJ}, and the resulting spacetime
cannot be directly handled by the ``pocket-based'' measure we employ
here.}, \beqa  \kappa_{ji} &=& \kappa_{ij}
\left(\frac{H_j}{H_i}\right)^4 \exp \left[-24
\pi^2\left(\frac{1}{\Lambda_i}-\frac{1}{\Lambda_j}\right)\right] \nonumber \\
&=&\kappa_{ij}\left(\frac{\Lambda_{j}}{\Lambda_i}\right)^2 \exp
\left[-24
\pi^2\left(\frac{1}{\Lambda_i}-\frac{1}{\Lambda_j}\right)\right]
\label{eqn:updown} \eeqa

Given the entire set of rates $\kappa_{ij}$, we can in principle
compute the bubble abundance $p_{\alpha}$ for each vacuum $\alpha$,
following the methods of Refs.\ \cite{GSPVW,SPV}.  An exact
calculation would require diagonalizing an $N\times N$ matrix.  But
as in Ref.\ \cite{SPV}, we can make the approximation that all
upward transition rates are tiny compared to all downward transition
rates from a given vacuum.

Once we have identified the dominant vacuum, the probability for any
vacuum $\alpha$ is given by (see the next section) \be
\label{eqn:pl} p_\alpha = \sum \frac{\kappa_{\alpha
a}\kappa_{ab}\cdots \kappa_{z*}} {(D_a-D_*)(D_b-D_*)\cdots(D_z-D_*)}
\ee where the sum is taken over all chains of intermediate vacua
$a,b,\ldots,z$ that connect the vacuum $\alpha$ to the dominant
vacuum.

\subsection{Bubble abundances by perturbation theory} As shown in
Ref.~\cite{GSPVW}, the calculation of bubble abundances $p_j$
reduces to finding the smallest eigenvalue $q$ and the corresponding
eigenvector $s$ for a huge $N\times N$ recycling transition matrix
$\mathbf{R}$.  Bubble abundances are given by \be p_j\propto
\sum_\alpha H_\alpha^q \kappa_{j\alpha} s_\alpha \approx \sum_\alpha
\kappa_{j\alpha} s_\alpha. \label{pJaume} \ee where the summation is
over all recyclable vacua which can directly tunnel to $j$.
$H_\alpha$ is the Hubble expansion rate in vacuum $\alpha$ and we
take $H_\alpha^q \approx 1$ because $q$ is an exponentially small
number.

In a realistic model, we expect $N$ to be very large. In the numerical
example of Ref.~\cite{SPV} $N\sim 10^7$, while for a realistic string
theory landscape we expect $N \sim 10^{500}$
~\cite{Susskind,Douglas,AshokDouglas,DenefDouglas}. Solving for the
dominant eigenvector for such huge matrices is numerically impossible.
However, in Ref.~\cite{SPV,SPthesis} the eigenvalue problem was solved via
perturbation theory, with the upward transition rates (see
Eq.~(\ref{eqn:updown})) playing the role of small expansion
parameters. Here we extend this procedure to all orders of
perturbation theory.

We represent our transition matrix as a sum of an unperturbed matrix
and a small correction,
\be
\mathbf{R}=\mathbf{R^{(0)}}+
\mathbf{R^{(1)}},
\ee
where $\mathbf{R^{(0)}}$ contains all the
downward transition rates and $\mathbf{R^{(1)}}$ contains all the
upward transition rates.  We will solve for the zero'th order
dominant eigensystem $\{q^{(0)},\mathbf{s^{(0)}}\}$ from
$\mathbf{R^{(0)}}$ and then include contributions from
$\mathbf{R^{(1)}}$ to all orders of perturbation theory.

If the vacua are arranged in the order of increasing $\Lambda$, so
that
\be
\Lambda_1 \leq \Lambda_2 \leq \ldots \leq \Lambda_N,
\ee
then $\mathbf{R^{(0)}}$ is an upper triangular matrix.  Its
eigenvalues are simply equal to its diagonal elements,
\be
R^{(0)}_{\alpha\alpha} =-\sum_{j<\alpha}\kappa_{j\alpha} \equiv
-D_\alpha. \label{Ralpha}
\ee
Hence, the magnitude of the smallest
zeroth-order eigenvalue is
\be
q^{(0)}=D_{{*}} \equiv {\rm
min}\{D_\alpha\}. \label{q0}
\ee

Downward transitions from ${*}$ will bring us to the
negative-$\Lambda$ territory of terminal vacua \cite{SPV}.  Terminal
vacua do not belong in the matrix $\mathbf{R}$; hence,
$R_{\beta{*}}=0$ for $\beta\neq{*}$, and we see that the zeroth order
eigenvector has a single nonzero component,
\be
s^{(0)}_\alpha=
\delta_{\alpha{*}}. \label{s0}
\ee
Thus, in fact, we could have
included only the diagonal elements of $\mathbf{R^{(0)}}$ and still
obtained the correct $q^{(0)}$ and $\mathbf{s^{(0)}}$.

Now we would like to include the effect of the upward transition rates
in the lower triangular matrix $\mathbf{R^{(0)}}$, to any given order
of perturbation theory.  To organize the calculation, we note that
the eigenvalues of an upper triangular matrix are just the diagonal
elements, and the eigenvectors are given exactly by the perturbation
series, which terminates after $N$ terms.  Thus we will take
just the diagonal elements of $\mathbf{R^{(0)}}$ as our unperturbed
matrix, and consider the rest of $\mathbf{R^{(0)}}$ and
$\mathbf{R^{(1)}}$ as a perturbation.  By summing all terms in the
perturbation series for this perturbation that involve $n$ elements of
$\mathbf{R^{(1)}}$, we find the perturbation term of order $n$ in the
small upward jump rates.

The procedure is the same as that involved in finding an eigenstate of
the Schr\"odinger equation in $n$th-order perturbation theory, working
in a basis of wavefunctions which diagonalize the unperturbed
equation.  See for example Eq.~(9.1.16) of Ref.~\cite{MorseFeshbach}.
Including all orders of perturbation theory, we find the result
\be
\label{eqn:pfinal}
s_a = \sum_{b=1}^N\cdots\sum_{z=1}^N
\frac{\kappa_{ab}}{(D_a-D_*)}
\cdots \frac{\kappa_{z*}}{(D_z-D_*)}
\ee
where there can be any number of terms in the sum and the vacuum * is
not summed over.   Combining  Eqs.\ (\ref{eqn:pfinal}) and
(\ref{pJaume}) gives Eq.\ (\ref{eqn:pl}).

\section*{Acknowledgments}
D. S.-P. was supported in part by grant RFP1-06-028 from The
Foundational Questions Institute (fqxi.org). I would like to thank
Alex Vilenkin and Ken Olum for many useful discussions and
suggestions during the course of this work.  Also thanks to
Jose-Juan Blanco-Pillado and Michael Salem for useful discussions.

\end{document}